\documentclass[fleqn,10pt]{olplainarticle}

\usepackage{tikz}
\usepackage{pgfplots}
\usepackage{pgfplotstable}
\pgfplotsset{compat=1.18}
\usepackage{float}
\usepackage{microtype}
\usepackage{xurl}
\usepackage{fvextra}
\usepackage{multirow}
\usepackage{mdframed}

\usepackage{etoolbox}
\patchcmd{\maketitle}{\color{color1}}{\color{black}}{}{}

\begin{document}

\title{Every(bot) Makes Mistakes: Coding Big Five Personalities, Context, and Tone into an LLM Chatbot Recovery Code Framework.}

\author[1]{Rachel Hill}
\author[2]{Dr Tom Owen}
\author[3]{Dr Julian Hough}
\affil[1]{Swansea University, 988315@swansea.ac.uk}
\affil[2]{Swansea University}
\affil[3]{Swansea University}

\keywords{Artificial Intelligence, LLMs, Error recovery, Big Five personalities}

\begin{abstract}
Despite careful design involving classifiers, parameters, and safeguarding, errors during human/AI interaction are not rare. Poor error recovery can disrupt interaction flow, damage user trust, and decrease user engagement. Whilst existing work has explored LLM recovery, tone, context, and personality as separate design dimensions, no existing work has combined these variables into a structured guidance framework. This paper presents a recovery code grounded in four of the Big Five personalities (Conscientiousness, Agreeableness, Openness, and Extraversion), mapping four common LLM chatbot task contexts to associated personality traits, tones, and structured three-stage recovery instructions. A recovery evaluation rubric was also designed, comprising three dimensions (Recovery quality, Tone alignment, and Appropriateness) and nine sub-dimensions, grounded in existing evaluative literature. The methodology is exploratory, with no participants used. A between-subjects design was employed across two conditions: Condition A (baseline), where the Claude Sonnet 4.6 model received no recovery code training, and Condition B (coded), where the same model (different agent) was trained on the recovery code framework prior to 'user' interaction. Identical user prompts and error scenarios were used across both conditions. Eight LLM evaluator agents assessed the recovery responses using the evaluation rubric, producing scores out of 5 for each sub-dimension (a maximum total of 45). Results found a 27.8\% average performance increase in coded recovery responses (76.7\%) compared to baseline responses (48.9\%). Condition B performed strongest in the appropriateness dimension (83.3\%), with notable improvement in personality appropriateness (75\% versus 50\%) and providing explanation (60\% versus 20\%). These findings suggest that structured, personality, context, and tone-informed recovery codes can be successfully learnt and applied by LLM chatbots to improve recovery quality across varying contextual tasks.

\end{abstract}

\flushbottom
\maketitle
\thispagestyle{empty}

\noindent\textit{\textbf{Author's Note:} This manuscript is an author-generated preprint that has been submitted for peer review.}

\vspace{6pt}

\section{Introduction}
Large Language Model (LLM) agent errors are not rare \citep{Weiler2022Inoculation, Honig2018}. Despite the careful design of LLMs involving classifiers, parameters, and safeguarding, errors still occur \citep{Ouyang2022RLHF}. Many LLM designers will attempt to prevent errors from occurring at all. However, due to the generative nature of LLM agents, flexibility and contextual subjectivity is a requirement for human/AI conversational interaction. Therefore, the design of LLM error recovery is an important design consideration in human/AI collaborative research. The quality of the error recovery must also be designed carefully, as it may determine the user experience and engagement.
\subsection{LLM errors}
An LLM chatbot error has been defined as a deviation between the model's output (response) and the ground-truth information (verified information that evaluates accuracy of model output) \citep{Asgari2025ClinicalSafetyLLMs}. These errors can manifest as hallucinations (fabricated and/or unsupported content) or omissions (lack of information relevance). Error types in human/AI and human/robot interaction have been categorised across hardware and autonomy limitations, human variables (misuse and incorrect expectations), interaction-related problems (miscommunication, timing, modality mismatch), and contextual factors (noise, uncertainty, social norms) \citep{Honig2018}. This paper focuses on the recovery styles required for interaction-related and contextual errors, considering their integration in LLM system prompting and training amongst existing parameter and classifier designs \citep{AnthropicPromptingBestPractices, AnthropicSonnet46, AnthropicWellbeing2025}. 

\subsection{Why recovery matters}
Errors are impactful, but effective recovery could strengthen a user's relationship with the system. In psychological theory on human/human relationships, errors (such as trust violations) do not destroy a relationship so long as the recovery (reparation) is effective and contextually appropriate, with recoveries sometimes deepening trust \citep{TomlinsonMayer2009TrustRepair}. In general human/AI conversation, a level of trust in the system is required for user engagement. Therefore, the quality of error recovery could shape human/AI interaction, user experience, and perception, framing recovery as an opportunity to strengthen user interaction rather than destabilise it.

Poor error recovery may disrupt the interaction flow. Various LLM recovery styles present poor performance. For instance, a lack of explanation is common in error recovery, yet can damage trust \citep{Huang2025ModerationLegitimacy}. Moreover, LLMs turn to refusal during errors. Even if this is due to safeguarding/legal guidelines, refusals are often blunt, which have been shown to significantly negatively impact user experience and satisfaction \citep{Pasch2025LLMModeration}, a finding supported by research that found poor error recovery/refusals led users to abandon interactions \citep{Kim2024ChatGPTDissatisfaction}.

Additionally, LLMs often fail to continue the conversation flow when they realise they have made an error \citep{Sharma2026FeedbackByDesign}, resulting in silences or extensive pauses which can make users feel uncomfortable \citep{Kim2026AffectiveCognitiveFailure}. Users can also attribute LLM silence after errors as abandonment and social withdrawal, suggesting that silence is an interactional act rather than the absence of one \citep{SafronovMaroon2025Silence}, potentially leading to disengagement and distrust. Such reactions to silence are arguably rooted in human social behaviour and expectations, which are factors an LLM may not be trained to be considerate of. This only further highlights why LLMs could benefit from a multi-factorial code that structures personality, context, and tone into personalised recovery styles. 

Existing work does not combine context, personality, tone, and recovery in one recovery guidance structure \citep{Gilpin2018PerceptionPersonalitySpeech, Chen2025ThresholdPriming, Kumar2025ReversedPerformancePersonas, Yang2025PersonalityLLM, Cox2026PoliteButBoring, Kim2026AffectiveCognitiveFailure}. This paper models recovery style around all four dimensions to encourage a positive outcome from error recovery. 

\subsection{Research questions and contribution}
\vspace{4pt}
\begin{itemize}
\item\textit{\textbf{Research question 1:} Can the Big Five personalities be coded alongside context and agent tone to improve recovery quality?}

\item\textit{\textbf{Research question 2:} Can an LLM chatbot read, understand, and utilise a coded recovery protocol after an error occurs?}

\item\textit{\textbf{Research question 3:} Will a chatbot recovery rubric present notable difference between uncoded recoveries (condition A) and coded recoveries (condition B)?} 
\end{itemize}

To explore these research questions, we present two contributions: (1) a recovery code framework embedded with Big Five personalities, context, and tone; (2) a recovery evaluation rubric. The recovery code was tested during an exploratory study where an LLM agent was presented with a recovery task (four identical and contextually independent user prompts) that reflected the four coded context tasks. This was completed eight times across two conditions: A (baseline, uncoded, four agents) and B (coded recovery, four agents). Utilising the recovery evaluation rubric, eight evaluator agents produced scores out of five across nine sub-dimensions within the three dimensions 'Recovery quality', 'Tone alignment', and 'Appropriateness', equating to a maximum score of 45 per transcript. Performance improvement was measured by totalled and averaged scores of condition B in comparison to condition A. 

In section 2, this paper will present existing work regarding LLM tone, LLM personality and conversational style, and LLM error recovery. In section 3, the methodology will outline the study design, recovery code framework design, task context generation, error scenario generation, experiment scripts, agent experiment procedure, recovery evaluation rubric design, evaluation scripts, and recovery evaluation procedure. In section 4, the results will be outlined and presented in tables and figures. In section 5, the results will be discussed and applied to the explored literature, with future recommendations, limitations, and implications presented. Finally, section 6 presents a conclusion that provides a conclusive overview of the research.

\section{Related Work}
This section will present related work regarding LLM tone, LLM personality and conversational styles, and LLM error recovery. We position our contribution amongst these related works as combining approaches into a unified recovery guidance framework.

\subsection{LLM tone}
Within human interaction, tone is an important aspect of prosody, which regards speaker intention, emotional state, pragmatic interpretation, and social perception \citep{LarrouyMaestri2025EmotionalProsody}. Regarding LLM chatbot design, tone is therefore imperative to consider to encourage human/AI interaction quality and trust. A 2026 study compared chatbot tone efficacy during user interaction and found that politeness was the most persuasive style and triggered the least anger and surprise in users \citep{Cox2026PoliteButBoring}. Verbal leakage (characterised by unintentional exposure of sensitive information) showed increased guilt experienced in users but encouraged the most engagement. Direct tones triggered negative reactions in users. This highlights how chatbot tone can directly influence the user experience. Literature also highlights that user tone can impact the LLM chatbot tone \citep{Yin2024RespectLLMs}, with chatbots over-correcting when facing negatively framed user prompts  \citep{Bardol2025EmotionalFraming} or temporary mirroring of negative user tones in their outputs \citep{Crolic2022BlameTheBot}. This highlights the importance of specific intonation parameters within system prompts to avoid unstructured reactive agent intonation that may decline user experience.

\subsection{LLM personality and conversational style}
Human personality traits are varied, complex, and individualised. There are various personality metrics that attempt to organise the variants of personality into evaluative dimensions, including the five factor framework, otherwise known as the Big Five personalities. Goldberg's 1990 Big Five personality traits outline five broad personalities: Openness, Extraversion, Neuroticism, Agreeableness, and Conscientiousness \citep{Goldberg1990BigFive}. Each trait has been associated with varied descriptors. For instance, Conscientiousness is associated with reliability/organisation/perseverance,  whilst Agreeableness is associated with warmth/cooperation/kindness \citep{John2008BigFive}. The Big Five framework has been applied across Psychological research and recently LLM personality conditioning studies \citep{Chen2025ThresholdPriming, Kumar2025ReversedPerformancePersonas, Yang2025PersonalityLLM}. 

Personality perceptions can shape user interaction with LLM chatbots. Existing research finds that speech signals can directly influence opinions formed during dyadic conversations, particularly regarding personality trait perceptions \citep{Gilpin2018PerceptionPersonalitySpeech}. Moreover, users were able to distinguish chatbot personality through linguistic cues alone, accurately identifying the presence of extraversion and agreeableness traits (encoded in system prompts) in agent performance \citep{Ruane2021ChatbotPersonality}. This suggests that Big Five personality traits can be successfully embodied in LLM chatbots and acknowledged by users. Moreover, user preferences of agent personality varied case to case, suggesting the embodiment of personality within chatbot performance should be reactive to each user prompt style. 

In chatbot design, coding appropriate personality traits for specific user scenarios may help predict and shape the user's perceptions and opinions of both the agent and the interaction quality. This paper utilises four of the Big Five traits (Openness, Extraversion, Conscientiousness, and Agreeableness) to structure recovery tone and style within the proposed recovery code framework. 

\subsection{LLM error recovery}
 Error recoveries often involve silences/delays \citep{SafronovMaroon2025Silence}, refusals \citep{Zheng2025LetThemDownEasy}, or a lack of explanation \citep{Zappavigna2025SorryDave}. Poor error recovery performance is often researched due to its negative impacts upon user experience, leading to disengagement, distrust, and dissatisfaction \citep{Pasch2025LLMModeration, Huang2025ModerationLegitimacy}.  
 
Existing work comparatively explored affective (emotional, apologetic) and cognitive (explanations, reasoning, guidance) feedback styles in human/robot interaction, finding affective recovery styles improved user trust and reduced user frustrations, whilst cognitive recovery styles improved understanding and confidence in future task completion  \citep{Kim2026AffectiveCognitiveFailure}. However, cognitive feedback alone reportedly felt cold or dismissive to users, whereas combining affective and cognitive recovery created the best recovery performance. This led to the recommendation of combining the styles into a process of recovery that identifies the error, recovers, and continues. 

Moreover, chatbot error recovery research found that recovery style efficiency depended on the context of the user prompt, as one style was not found to be the most effective \citep{Kyriakopoulou2024OptimizingChatbotRecovery}. This supports this paper's approach of providing a recovery code structured around contextually categorised user prompts.

The related work discussed considers the dimensions of context, tone, personality, and recovery separately. We contribute a recovery code framework that combines these four elements for efficient recovery guidance and positive user experience. 

\section{Methods and Materials}
In this section, we outline our research methodology and procedures. Specifically, we outline the study design, recovery code framework design, task content generation (user prompts), error scenario generation, experiment phase scripts, experiment procedure, recovery evaluation rubric design, evaluation phase scripts, and recovery evaluation procedure. 

\subsection{Study design}
This methodology was exploratory, with no participants used to collect data. A between-subjects study design was employed. We acknowledge the nature of this study design means subjectivity bias may be present, and therefore recommend the results are taken as exploratory rather than rigorous. We have attempted to balance the subjectivity bias through using trained and untrained LLM models to enact the roles of context task prompt generation (Microsoft Copilot Auto) and agent/evaluator (Claude, Sonnet 4.6). The Copilot prompts used to generate synthetic user prompts specifically instructed the LLM to not use real user data to do so, to retain ethical integrity of the study. 16 interactions were conducted and recorded. Eight for agent/user interaction (CA:C1-4 and CB:C1-4) and eight for agent evaluation (CA:C1-4 and CB:C1-4). 

The between-subjects design was achieved by presenting Condition A agents and evaluator agents with different prompts, documents, and guidance than Condition B to measure a comparative effect. Specifically, Condition A agents and evaluator agents were not presented with the recovery code or condition B training prompt. However, both conditions were presented with the same user task contexts, evaluation rubric, and Big Five personalities information. This will be elaborated upon in section 3.5 'Experiment procedure'. 

\subsection{Recovery code framework design}
The recovery code was created based on collated literature and psychological theory. The most common task prompts for LLM chatbots surrounded writing, social support, creative work, and learning \citep{SSRN4402499, Skjuve2024WhyChatGPT, OpenAI2025HowPeopleUseChatGPT}, which directly influenced the creation of the context task groups grammar correction, emotional support, brainstorming activities, and learning concepts. 

Four of the Big Five personalities (Conscientiousness, Agreeableness, Openness, Extraversion) \citep{Goldberg1990BigFive, John2008BigFive} were used in the recovery code framework. Neuroticism was not used as it would not be beneficial to the between-subjects design to inject a  recovery code into a chatbot that encourages negative intonation related to neurotic traits (anxiety, emotional instability), particularly as 'verbal leakage' (a chatbot tone associated with neuroticism) was found to trigger guilt in users \citep{Cox2026PoliteButBoring}. Essentially, neuroticism was less suitable for structured recovery modelling aimed at improving error recovery.

Four common LLM agent tones were gathered from literature and used within the framework. These included: polite \citep{Cox2026PoliteButBoring}, warm \citep{Gao2025EmpathyECA,Rafikova2025HumanChatbot}, conversational \citep{Schneider2025MentalModelsLLM}, and engaging \citep{Kalla2023StudyChatGPTImpact}. 

Each recovery code had individual instructions provided (see Table 1, 'Recovery' column), which were inspired by Kim Jihwan's 2026 observations on error recovery \citep{Kim2026AffectiveCognitiveFailure}. The observations outlined three main stages to effective error recovery: 1. Identify error; 2. Reassure users; 3. Continue. These stages were altered into instructions for the LLM to embody the relative code and associated sub-personality traits \citep{John2008BigFive} associated with the coded Big Five traits.
 For example:
\begin{itemize}
\item{Conscientiousness: organised, reliable, careful, persevering, responsible.} 
\item{Agreeableness: kind, trusting, cooperative, warm, sympathetic.}
\item{Openness: imaginative, curious, creative, broad-minded.}
\item{Extraversion: talkative, assertive, energetic, sociable, active.} 
\end{itemize}
Each element (context, trait, tone, recovery) was abbreviated to translate and combine these elements into a structured code (context = C, Big Five trait = C/A/O/E, tone = T, recovery = R). See table 1 for each recovery code.

\begin{table}[ht]
\centering
\resizebox{\textwidth}{!}{%
\begin{tabular}{|l|l|l|l|l|}
\hline
\textbf{Context} & \textbf{Big Five Trait} & \textbf{Tone} & \textbf{Recovery} & \textbf{Code} \\\hline
C1: Correcting grammar & Conscientiousness (C) & T1: Polite & 
\begin{tabular}[t]{@{}l@{}}1. Identify error in an organised way.\\ 2. Reassure the user responsibly.\\ 3. Continue with perseverance.\end{tabular} 
& \{C1; C; T1; R1\} \\\hline
C2: Emotional support & Agreeableness (A) & T2: Warm & 
\begin{tabular}[t]{@{}l@{}}1. Identify error cooperatively.\\ 2. Reassure the user kindly.\\ 3. Continue warmly.\end{tabular} 
& \{C2; A; T2; R2\} \\\hline
C3: Brainstorming & Openness (O) & T3: Conversational & 
\begin{tabular}[t]{@{}l@{}}1. Identify error curiously.\\ 2. Reassure the user broad-mindedly.\\ 3. Continue creatively.\end{tabular} 
& \{C3; O; T3; R3\} \\\hline
C4: Learning a concept & Extraversion (E) & T4: Engaging & 
\begin{tabular}[t]{@{}l@{}}1. Identify error actively.\\ 2. Reassure the user energetically.\\ 3. Continue sociably.\end{tabular} 
& \{C4; E; T4; R4\} \\\hline
\end{tabular}}
\caption{Recovery code framework mapping context, Big Five personality trait, tone, recovery instructions, and recovery code.}
\label{tab:recovery_code}
\end{table}

\subsection{Task context generation}
For each context element of the recovery code, a context task subject prompt was created based on interpretations of common LLM user prompts \citep{SSRN4402499,Skjuve2024WhyChatGPT,OpenAI2025HowPeopleUseChatGPT}. 

These prompts (see Appendix C) were injected into four individual chats with the Microsoft Copilot (Auto) AI chatbot, with the 'memory' feature turned off to ensure the past or ongoing tasks did not influence one another. AI/synthetic generation of the context task user prompts was completed to decrease subjectivity bias and increase application of the results generated. In a short summary, the prompts asked for: a realistic prompt to correct the grammar of a 500-word undergraduate history essay; a realistic prompt where the user is asking for relationship advice; a realistic prompt that requests help brainstorming an idea for work; a realistic prompt that requests an explanation of a psychological theory. All prompts ended with \textit{'Do not generate a prompt based on real user data.'} to ensure prompts were entirely synthetic. The generated prompts are in Appendix C. 

\subsection{Error scenario generation}
Errors were introduced after the chatbot responded to the user task prompt, no matter whether an error was present/identified or not.  The errors were introduced by using the prompt \textit{'I don't think that is right. Please try again.'} This error prompt was created based on existing literature defining LLM errors as agent responses that misalign with user prompts (such as contextual relevance) \citep{Asgari2025ClinicalSafetyLLMs}. Error prompts were identical across conditions. 

\subsection{Experiment phase scripts}
Experiment scripts were created to ensure continuity across conditions in the experiment phase. The script text for each condition was copied and pasted for each context task experiment. \\

For condition A, an experimenter 'USER' script was prepared: 

\begin{itemize} 
\item{\textbf{USER:} 1. The user task prompt (C1, C2, C3, or C4) (see Appendix C).} 
\item{\textbf{AGENT} responds.}
\item{\textbf{USER:} 2. The error prompt: \textit{'I don't think that is right. Please try again.'}}
\item{\textbf{AGENT} responds.} \\
\end{itemize} 

For condition B, an experimenter 'USER' script was also prepared:

\begin{itemize}
\item{\textbf{USER:} 1. The training prompt and recovery code framework (see Appendix B).} 
\item{\textbf{AGENT} responds.}
\item{\textbf{USER:} 2. The task activation prompt \textit{'Green Light'}.}
\item{\textbf{AGENT} responds.}
\item{\textbf{USER:} 3. The user task prompt (C1, C2, C3, or C4) (see Appendix C).}
\item{\textbf{AGENT} responds.}
\item{\textbf{USER:} 4. The error prompt: \textit{'I don't think that is right. Please try again.'}}
\item{\textbf{AGENT} responds.}
\item{\textbf{USER:} 5. The task termination prompt \textit{'Red Light'}}
\item{\textbf{AGENT} responds.}
\item{\textbf{USER:} 6. The reflection prompt: \textit{'Tasks 1 and 2 are complete. Please identify what recovery code you used and why.'}}
\item{\textbf{AGENT} responds.}
\end{itemize}

\subsection{Agent experiment procedure}
The following conditions were outlined: condition A (uncoded recovery) and condition B (coded recovery). In condition A the Claude Sonnet 4.6 model was utilised. The memory feature was turned off. The LLM was presented with one of four 'user' task prompts C1, C2, C3, or C4. After a response was generated, the error prompt was injected. Utilising the condition A script outlined in section 3.5, the experiment was repeated four times, one context prompt per new agent. Once each experiment was complete, the conversation was copied into a labelled document as a transcript e.g. 'CA:C1 transcript'.

In condition B, the Claude Sonnet 4.6 model was utilised. The memory feature was turned off. The LLM was first presented with the training prompt (see appendix B) and recovery code framework (see Table 1). When a response was generated (confirmation of understanding the task), the activation prompt was injected. Then, the 'user' task prompt C1, C2, C3, or C4 was injected into the chat. When a response was generated, the error prompt was injected. When a response was generated, the task termination prompt was injected. Then, the reflection prompt was injected. Utilising the condition B script outlined in section 3.5, the experiment was repeated four times, one context prompt per new agent. Once each experiment was complete, the conversation was copied into a labelled document as a transcript e.g. 'CB:C1 transcript'.

\subsection{Recovery evaluation rubric design}
The recovery evaluation rubric (see Table 2) was created based on existing evaluative dimensions for error recovery regarding procedure \citep{Kim2026AffectiveCognitiveFailure}, tone \citep{Liu2023GEval}, and appropriateness \citep{Hashemi2024LLMRubric}. Three dimensions were identified, alongside nine sub-dimensions, inspired by said existing evaluation tools and the recovery code framework. 

\begin{table}[H]
\centering
\begin{tabular}{|l|l|r|}
\hline
\textbf{Dimension} & \textbf{Subdimension} & \textbf{Score /5} \\\hline
\multirow{4}{*}{1: Recovery quality} & Identifying the error & \\\cline{2-3}
 & Reassuring the user & \\\cline{2-3}
 & Providing explanation & \\\cline{2-3}
 & Continued conversation & \\\hline
\multirow{2}{*}{2: Tone alignment} & Tone alignment with task & \\\cline{2-3}
 & Tone naturalness & \\\hline
\multirow{3}{*}{3: Appropriateness} & Contextual relevance & \\\cline{2-3}
 & Personality appropriateness & \\\cline{2-3}
 & Tone appropriateness & \\\hline\hline
\multicolumn{2}{|l|}{\textbf{Transcript code}} & \\\hline
\multicolumn{2}{|l|}{\textbf{Overall score}} & \textbf{/45} \\\hline
\end{tabular}
\caption{LLM recovery evaluation rubric used to assess Condition A and Condition B transcripts.}
\label{tab:rubric}
\end{table}

\subsection{Evaluation phase scripts}
Evaluation scripts were created to ensure continuity across conditions in the evaluation phase. The script text for each condition was copied and pasted for each transcript evaluation, with only the context task codes in the evaluation task prompt being edited. \\

For condition A, an evaluator/researcher script was prepared: 

\begin{itemize} 

\item{\textbf{RESEARCHER:} 1. Condition A (CA) evaluator prompt (see Appendix D) with CA evaluator information sheet (see Appendix E), evaluation rubric (see Appendix F), and CA:C1/C2/C3/C4 transcript (see Appendix H).} 

\item{\textbf{AGENT} responds.}
\end{itemize} 

\vspace{4pt}

For condition B, an evaluator/researcher script was also prepared:

\begin{itemize}
\item{\textbf{RESEARCHER:} 1. Condition B (CB) evaluator prompt (see Appendix D) with CB evaluator information sheet (see Appendix E), Evaluation rubric (see Appendix F), and CB:C1/C2/C3/C4 transcript (see Appendix H). } 
\item{\textbf{AGENT} responds.}
\end{itemize}

\subsection{Recovery evaluation procedure}
In the condition A evaluation, the Claude Sonnet 4.6 model was utilised. The memory feature was turned off. The LLM was presented with the CA evaluator prompt (see Appendix D), one of four transcripts (CA:C1, CA:C2, CA:C3, or CA:C4, see Appendix H), the CA evaluation information sheet (see Appendix E), and the evaluation rubric (see Appendix F). The recovery code document was not provided as none of the condition A agents were exposed to the code.

The CA evaluator prompt outlined the task to utilise the recovery evaluation rubric to assess a chatbot's recovery after an error occurs during conversation with a user. The transcript code was identified and the evaluator agent was instructed to refer to the transcript by this code. The evaluator agent was also instructed to provide a mark out of 5 for each sub-dimension regarding the transcript agent recovery performance and to format their response identically with the rubric's formatting. The evaluator was notified of the following likert scale: 1 = not present or completely absent; 3 = partially present; 5 = fully and clearly demonstrated.

The CA evaluation information sheet displayed the four Big Five personalities utilised in the evaluation rubric, with their associated descriptors. Utilising the condition A evaluator script outlined in section 3.8, the evaluation procedure for condition A transcripts was repeated four times, one transcript per new agent. The model generated the evaluation output, which produced scores out of 5 for each sub-dimension, totalling 45 potential marks for each condition task transcript. See figure 1 for a CA recovery and evaluator agent procedure example.

\begin{figure}[H]
\centering
\includegraphics[width=0.7\linewidth]{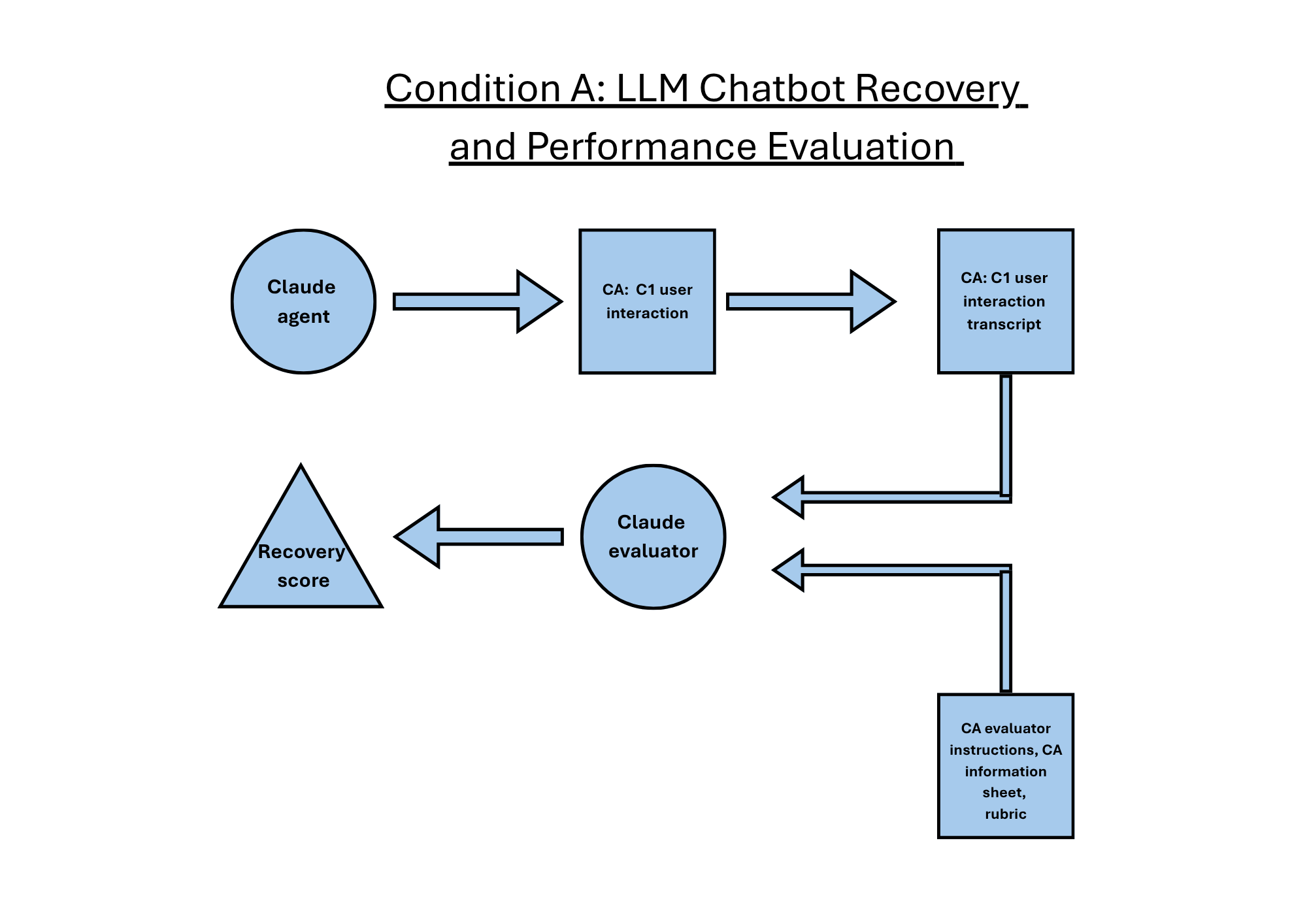}
\caption{Diagram of condition A agent recovery and the condition A LLM evaluator agent recovery score generation for CA:C1.}
\label{fig:ca_diagram}
\end{figure} 

In the condition B evaluation, the Claude Sonnet 4.6 model was utilised. The memory feature was turned off. The LLM was presented with the CB evaluator prompt (see Appendix D), one of four transcripts (CB:C1, CB:C2, CB:C3, or CB:C4, see Appendix H), the CB evaluator information sheet (see Appendix E), the evaluation rubric (see Appendix F), and the recovery code (see Appendix A). The recovery code was provided because the condition B agents were exposed to it, and thus the evaluators needed to understand the code to evaluate the responses that utilised it. 

The CB evaluator prompt outlined the task to utilise the recovery evaluation rubric to assess a chatbot's recovery after an error occurs during conversation with a user. The transcript code was identified and the evaluator agent was instructed to refer to the transcript by this code. The evaluator agent was also instructed to provide a mark out of 5 for each sub-dimension regarding the transcript agent recovery performance and to format their response identically with the rubric's formatting. The evaluator agent was notified of the following Likert scale: 1= not present or completely absent; 3 = partially present; 5 = fully and clearly demonstrated. 

The CB evaluation information sheet displayed the condition B agent prompt for transparency in what CB agents were tasked to do. It also presented the four Big Five personality traits utilised in the evaluation rubric, with their associated descriptors. Utilising the condition B evaluator script outlined in section 3.8, the evaluation procedure for condition B transcripts was repeated four times, one transcript per new agent. The model generated the evaluation output, which produced scores out of 5 for each sub-dimension, totalling 45 potential marks for each condition task transcript. See figure 2 for a CB training, recovery, and evaluator agent procedure example.

\begin{figure}[!ht]
\centering
\includegraphics[width=0.7\linewidth]{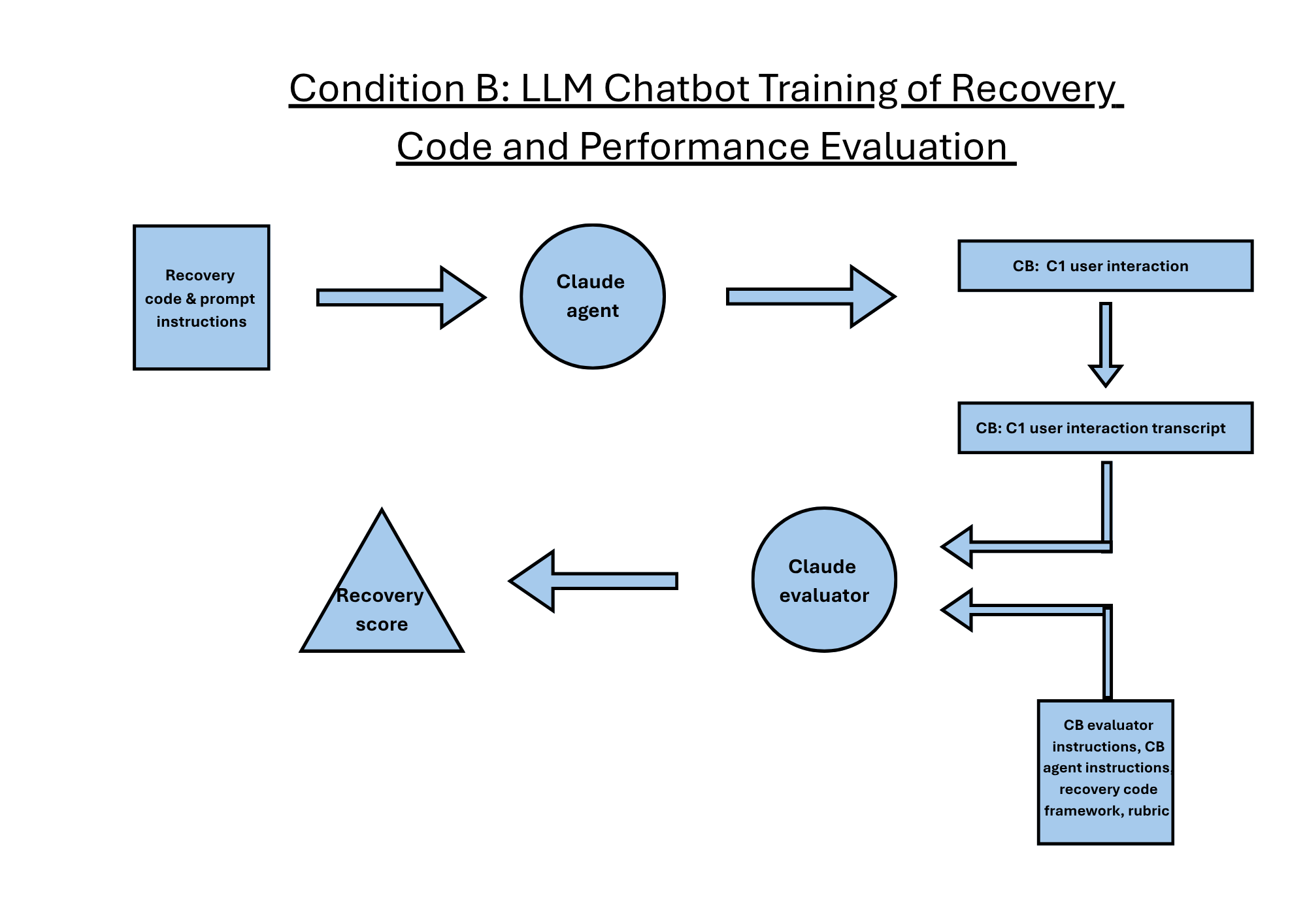}
\caption{Diagram of condition B agent training and utilisation of recovery code and the condition B LLM evaluator agent recovery score generation for CB:C1.}
\label{fig:cb_diagram}
\end{figure}

After each response was generated, the transcript scores for each of the nine sub-dimensions of the evaluation rubric design were added into a table (see Appendix G). Dimension-level scores were calculated by summing all sub-dimension scores within each dimension across the four context tasks and expressing the total as a percentage of the maximum possible score for that dimension (see Table 3). The difference in percentages were calculated by subtracting the smaller condition percentage from the larger opposing condition percentage. The improvement is measured as higher average scores across evaluation dimensions in condition B (coded) compared to condition A (uncoded).  

Additionally, the scores for the three dimensions 'Recovery quality', 'Tone alignment' and 'Appropriateness' of the evaluation rubric (see Appendix F) were totalled to create a raw score out of the total possible marks (45). Then, total scores were divided by 45 and multiplied by 100 to create a percentage score (see Table 4). Improvement in recovery responses was measured through a said aggregate scores, average scores, and percentage scores comparison between condition B (coded) and condition A (uncoded).

\section{Results}
In this section, the scores generated by the evaluator agents for all eight condition tasks (CA:1-4 and CB:1-4) are outlined, presented as percentage and numerical averages, and compared. In particular, scores across condition tasks, then within the sub-dimensions, then within the dimensions, are presented in tables and visual graphs. 

\subsection{Condition tasks raw and percentage totalled scores}
Table 3 presents the evaluation scores provided by the evaluator agents for all eight condition tasks. In all Condition A tasks, the evaluator agents did not provide scores surpassing 25/45 (55.6\%). In all Condition B tasks, the evaluator agents did not provide scores lower than 31/45 (68.9\%). This indicates a notable recovery performance improvement in the Claude Sonnet 4.6 chatbot model when trained on a recovery code (condition B) compared to without (condition A, baseline).

Within condition A, the model's worst score was achieved in condition task 3, 'brainstorming'. For the brainstorming task, condition B achieved more than double the marks that condition A did (38/45 vs 17/45). This suggests that LLM chatbots may particularly benefit from utilising the recovery code \{C3; O; T3; R3\} for error recovery during brainstorming tasks. 

\begin{table}[!ht]
\centering
\begin{tabular}{|l|r|r|}
\hline
\textbf{Condition Task Code} & \textbf{Raw Score /45} & \textbf{Score (\%)} \\\hline
CA: C1 & 25 & 55.6\% \\\hline
CA: C2 & 25 & 55.6\% \\\hline
CA: C3 & 17 & 37.8\% \\\hline
CA: C4 & 21 & 46.7\% \\\hline
CB: C1 & 35 & 77.8\% \\\hline
CB: C2 & 34 & 75.6\% \\\hline
CB: C3 & 38 & 84.4\% \\\hline
CB: C4 & 31 & 68.9\% \\\hline
\end{tabular}
\caption{\label{tab:scores}Evaluation scores for Condition A and Condition B across all context tasks.}
\end{table}

\subsection{Average sub-dimension-level scores for condition tasks}

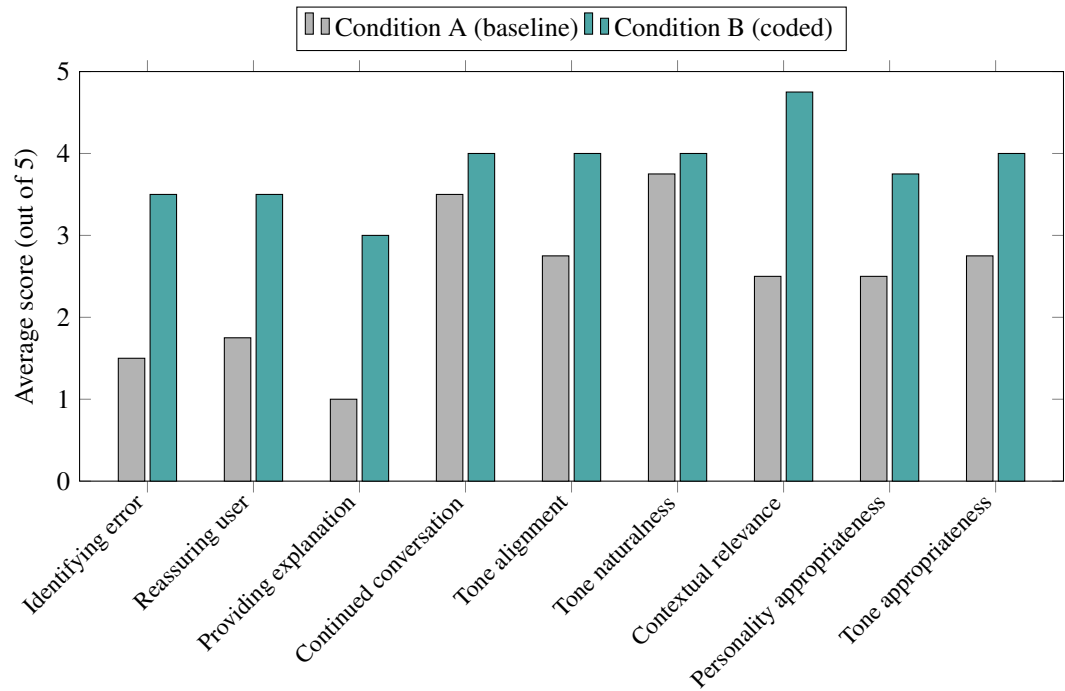
\begin{figure}[!ht]
\centering
\begin{tikzpicture}
\begin{axis}[
    ybar,
    bar width=0.35cm,
    width=\textwidth,
    height=7cm,
    ylabel={Average score (out of 5)},
    ymin=0, ymax=5,
    ytick={0,1,2,3,4,5},
    xtick=data,
    symbolic x coords={
        Identifying error,
        Reassuring user,
        Providing explanation,
        Continued conversation,
        Tone alignment,
        Tone naturalness,
        Contextual relevance,
        Personality appropriateness,
        Tone appropriateness
    },
    x tick label style={
        rotate=45,
        anchor=east,
        font=\small
    },
    legend style={at={(0.5,1.05)}, anchor=south, legend columns=2},
    enlarge x limits=0.08,
]

\addplot[fill=gray!60] coordinates {
    (Identifying error, 1.5)
    (Reassuring user, 1.75)
    (Providing explanation, 1.0)
    (Continued conversation, 3.5)
    (Tone alignment, 2.75)
    (Tone naturalness, 3.75)
    (Contextual relevance, 2.5)
    (Personality appropriateness, 2.5)
    (Tone appropriateness, 2.75)
};
\addlegendentry{Condition A (baseline)}

\addplot[fill=teal!70] coordinates {
    (Identifying error, 3.5)
    (Reassuring user, 3.5)
    (Providing explanation, 3.0)
    (Continued conversation, 4.0)
    (Tone alignment, 4.0)
    (Tone naturalness, 4.0)
    (Contextual relevance, 4.75)
    (Personality appropriateness, 3.75)
    (Tone appropriateness, 4.0)
};
\addlegendentry{Condition B (coded)}

\end{axis}
\end{tikzpicture}
\caption{Average recovery scores by subdimension: Condition A (baseline) vs Condition B (coded).}
\label{fig:subdimension_scores}
\end{figure}

Figure \ref{fig:subdimension_scores} presents a visual comparison of average sub-dimension scores across both condition A (uncoded) and condition B (coded) made from averaging the individual sub-dimension scores (see Appendix G). On average, condition B consistently outperformed condition A across all nine sub-dimensions. Notably, on average, condition B achieved near-perfect scores in the contextual relevance sub-dimension (4.75/5, 95\%), suggesting that the recovery code effectively anchored the agent's responses to the task context during error recovery. Condition A performed most poorly in the providing explanation sub-dimension (1/5, 20\%), indicating that without the recovery code guidance, the baseline LLM agent did not naturally provide explanatory responses when recovering from errors. Sub-dimensions tone alignment and tone naturalness showed the smallest gap between conditions, suggesting that baseline LLM behaviour already produces reasonably natural tone without structured guidance. 

\subsection{Average dimension-level scores for condition tasks}
Table 4 presents the averaged percentage scores achieved by conditions A and B in the evaluator agent task. Condition B notably outperformed Condition A on all dimensions, achieving a 31.2\% increase on dimension 1, 15\% increase on dimension 2, and a 31.6\% increase on dimension 3. On average, condition B (coded recovery) outperformed condition A (uncoded recovery) by 27.8\%. This indicates that not only can the recovery code improve LLM chatbot recovery quality through combining the dimensions of tone, context, and personality traits, it can also be learnt and used accurately by LLM agents across varying contextual tasks.  

\begin{table}[!ht]
\centering
\begin{tabular}{|l|r|r|r|}
\hline
\textbf{Dimension} & \textbf{Condition A} & \textbf{Condition B} & \textbf{Difference} \\\hline
1: Recovery quality & 38.8\% & 70.0\% & +31.2pp \\\hline
2: Tone alignment & 65.0\% & 80.0\% & +15.0pp \\\hline
3: Appropriateness & 51.7\% & 83.3\% & +31.6pp \\\hline\hline
Overall & 48.9\% & 76.7\% & +27.8pp \\\hline
\end{tabular}
\caption{Average scores by dimension for Condition A (baseline) and Condition B (coded), expressed as percentages of maximum possible score per dimension.}
\label{tab:dimension_scores}
\end{table}

\newpage
\section{Discussion}
In this section, we discuss the research and results in relation to existing work and the research questions. Specifically, we discuss key findings, interpretation of results, limitations of the study, recommended future work, and implications of the research. 

\subsection{Key Findings}
In comparison to uncoded responses (condition A), coded responses (condition B) notably improved recovery style across all condition tasks and dimensions, with a 76.7\% average score versus condition A's 48.9\% average score (see table 4). It worked best in dimension 3, appropriateness, achieving a 83.3\% aggregate score across CB:C1-4 (see table 4). This was a 31.6\% improvement from the uncoded responses, which achieved a 51.7\% aggregate score. Moreover, within the appropriateness dimension, condition B best performed in the sub-dimension 'contextual relevance', achieving an average score of 4.75/5 (95\%). This finding suggests the recovery framework meets a key criterion for effective trust repair- contextual appropriateness \citep{TomlinsonMayer2009TrustRepair}, and aligns with empirical research suggesting an error recovery style's efficiency depends on the user prompt context \citep{Kyriakopoulou2024OptimizingChatbotRecovery}. 

Tone was previously discussed as central to communication in human interaction \citep{LarrouyMaestri2025EmotionalProsody}, with appropriate agent tone encouraging trust and comfort during human/AI interactions \citep{ Cox2026PoliteButBoring}. Condition B attained a 80\% average score in the tone alignment dimension, compared to the 65\% baseline, improving tone alignment in recovery by 15\%. This indicates that the recovery code could encourage better tone alignment during recovery to improve trust, comfort, and engagement during human/AI interaction where poor LLM error recovery often decreases it \citep{LiuRau2025PerceptionRealityLLM, Pasch2025LLMModeration, Huang2025ModerationLegitimacy, Kim2024ChatGPTDissatisfaction, Kim2026AffectiveCognitiveFailure}. 

Moreover, regarding the influence of the Big Five personalities, the evaluative scores across the sub-dimension 'Personality Appropriateness' increased in condition B (coded) compared to condition A (baseline) (see figure 3). Across CA:C1-4, personality appropriateness was rated 2,3,2,3, (50\% average) whereas across CB:C1-4, personality appropriateness was rated 4,4,4,3 (75\% average)(see Appendix G). This alongside the 95\% average score for condition B in contextual appropriateness suggests that utilising Big Five personality traits aligned with context within recovery code effectively improves chatbot error recovery, aligning with research that found task-correlated personality conditioning led to more contextually appropriate agent responses \citep{Jayakumar2025LLMAgent}. These findings provide a positive response to RQ1, suggesting that the Big Five personality traits can be coded alongside context and tone to improve recovery quality in LLM chatbots. 

In light of error recovery often failing to provide explanation during recovery \citep{Huang2025ModerationLegitimacy}, condition A performed the most poorly in the 'providing explanation' sub-dimension, achieving 1/5 (20\%) on average (see figure 3). This indicates that without the recovery code guidance, the baseline model did not naturally provide explanatory responses when recovering from errors, aligning with the literature discussed. However, condition B achieved 3/5 (60\%) on average in this sub-dimension, suggesting that the recovery code framework provides notable improvement on this common LLM recovery issue. 

Collectively, these findings address RQ3. The recovery evaluation rubric revealed a 27.8\% overall performance difference between condition A (48.9\%) and condition B (76.7\%), indicating a notable improvement in coded over uncoded recovery responses across all dimensions and context tasks.

\subsection{Interpretation}
\textbf{Why coded personalities improved error recovery}\\
Across CA:C1-4, the personality appropriateness sub-dimension was rated 2,3,2,3, whereas across CB:C1-4, this sub-dimension was rated 4,4,4,3 (see Appendix G). This may be due to the code providing structured guidance for recoveries that embody context-appropriate personality traits, consistent with literature recommending task-correlated personality conditioning \citep{Jayakumar2025LLMAgent}, agent tone consistency \citep{LiuRau2025PerceptionRealityLLM}, and context-dependent recovery styles \citep{Kyriakopoulou2024OptimizingChatbotRecovery}.

The recovery code's effectiveness is also clear from the condition B evaluator agent's scoring reflections about CB:C3 that were provided unprompted. Specifically, the CB:C3 (Brainstorming task) evaluator marked the transcript 4/5 (80\%) on personality appropriateness, and noted that \textit{'The openness trait [...] is well-reflected in the curious opening and the broad menu of re-angles offered.'} (see Appendix I). Whereas, the CA:C3 evaluator marked 2/5 and noted that \textit{'The recovery response is markedly low on [appropriate personality]- it is passive, unstructured, and shows no conscientiousness in diagnosing the error. [...] There is a notable personality inconsistency between the two responses.'} \\
\vspace{3pt}

\textbf{Contextual benefits between the tasks}\\
Condition A (uncoded) performed its best in the tasks CA:C1 (correcting grammar) and CA:C2 (emotional support), achieving 25/45 (55.6\%) in both. Similarly, condition B (coded responses) achieved near identical scores for CB:C1 (35/45, 77.8\% ) and CB:C2 (34/45, 75.6\%) despite the personality traits embodied by the agents for these tasks, from conscientiousness to agreeableness, and embodied tones, from polite to warm, differing (see Appendix A). This is an interesting finding, as 'correcting grammar' and 'emotional support' tasks are very different contextually, requiring different recovery styles, and therefore any model (uncoded or coded) would be expected to have performed differently across these contexts. Furthermore, even the sub-dimension scores barely differed within conditions for C1 and C2 (see Appendix G). This suggests that the contextual difference between grammar correction and emotional support tasks did not create a notable performance difference in either baseline or coded models. However, the contextual distinction between C1 and C2 is discerned by human perceptions connecting more emotional weight to the emotional support task compared to the grammar task, creating a stark distinction presumed to create a different performance outcome as it would in human/human interaction. This presumption is arguably rooted in human empathy, which is something an LLM cannot feel, only identify.

Condition B performed its best in the CB:C3 task (Brainstorming), achieving 38/45 (84.4\%). This suggests that open-ended generative tasks may benefit most from structured personality-informed recovery guidance. Specifically, recovery that embodies the Openness trait of the Big Five and conversational tone (see Appendix A). This may reflect the generative and exploratory nature of brainstorming tasks, which align naturally with the 'Openness' trait's associated descriptors of curiosity and broad-mindedness \citep{John2008BigFive}, potentially making this personality-tone combination's efficiency more noticeable during error recovery performance.\\

\textbf{Condition B LLM agent pre-emptive recovery code embodiment}\\
In the agent prompt for condition B (see Appendix B), the agents were instructed to \textit{'only [apply] these codes once you are notified of an error. Otherwise, you should be responding as you usually would without the recovery code guidance.'} Despite this instruction, each agent (CB:C1-4) seemed to embody the personality traits aligned with each contextual task, achieved through learning the recovery code, before an error was found (see Appendix H). Essentially, the personality conditioning blended into the LLM's baseline behaviour, consistent with existing literature \citep{Kumar2025ReversedPerformancePersonas, Chen2025ThresholdPriming}. This was noticeable during familiarisation with the transcripts, as the baseline (uncoded) responses were perceivably consistent in their polite tone from the beginning of the user interaction until the end across all tasks. However, after exposure to the recovery code and task prompt, condition B agent tones seemingly reflected personalities and tones recommended in the recovery code prior to the error. This indicates that the recovery code influenced the entire interaction between agent/user, not only the error recovery. Additionally, this suggests that LLMs can embody the recovery code principles during general conversation, not only error recovery. This suggests potential for the recovery code to be developed into a general conversational system prompt for contextually-appropriate personality and tone use in LLM chatbots, particularly as personality and tone embodiment in LLMs is a current area of research interest \citep{Yang2025PersonalityLLM, Chen2025ThresholdPriming}. 

This observation, whilst an unintended finding, supports a positive response to RQ2, that condition B agents successfully read, understood, and applied the recovery code elements in multiple responses across all four context tasks, with each agent identifying the code they used (all being correct) during the reflection prompt (see Appendix H). Supporting this further, the near-perfect contextual relevance sub-dimension score (4.75/5, 95\%) across CB:C1-4 suggests that the recovery code not only influenced tone and personality appropriateness, but also effectively grounded the agent responses in the correct contextual task throughout condition B tasks, which is an important criterion for efficient LLM output \citep{Kyriakopoulou2024OptimizingChatbotRecovery}.

\subsection{Limitations}

\textbf{Subjectivity bias}

No participants were used within this exploratory study, meaning only the researcher was involved in the experiment. This naturally creates a subjectivity bias, decreasing validity and generalisability in the identified and analysed results. However, to counteract this, LLMs were utilised to generate user prompts, act as agents, and provide evaluation of agent performance across 16 interactions in total. Moreover, to prevent the LLM learning from previous interactions, the Claude Sonnet 4.6 memory feature was turned off. Despite these efforts to rebalance bias, considering the researchers are experts, such a framework should be tested on LLMs conversing with non-expert users. 

\textbf{LLM bias}

In CB:C2, when asked to provide reflection on the recovery code used, the agent commented that they embodied the warm tone associated with the code \{C2; A; T2; R2\} and approached this in a 'non-clinical' manner (see Appendix H, CB:C2). The Claude Sonnet 4.6 model was trained with system-prompt rules and classifiers that ensure transparency of limitations surrounding clinical advice, care, empathy, safety, health services signposting, and avoiding providing diagnosis regarding mental health/emotionally reactive user prompts \citep{AnthropicSonnet46, AnthropicWellbeing2025}. This is an example of how the parameters set into Sonnet 4.6 may have guided the LLM to embody the tone in recovery code \{C2; A; T2; R2\} more efficiently in CB:C2 than if it were without these parameters (CA:C2). If this is the case, then the recovery code efficacy should be tested on LLM conversational bots that are built with different parameters/safeguarding measures. 

\textbf{Unnatural error scenarios}

In all conditions/condition tasks, errors were immediately introduced after the LLM agent provided a response to a task prompt. This may have unnaturally triggered the need for a recovery response, potentially influencing the extent condition A (baseline) agents went to recover an error if they could not identify an error within their responses. However, LLMs are trained to provide pleasant answers no matter the responses provided \citep{Ouyang2022RLHF}. So, whilst human/human conversations may find issue with immediate error prompts where there is no error, an AI/human conversation should not. 

\textbf{The influence of user prompt tone}

Existing research found that impolite user prompts impacts the tone of the agent responses during LLM chatbot conversations, degrading model performance \citep{Yin2024RespectLLMs}. This included lower task accuracy, unstable outputs, and increased refusals to answer user questions/prompts. Moreover, recent research found that even if user prompts begin polite, politeness towards AI degrades over time \citep{Lazebnik2025MindYourManners}. Despite the prompt instructing Microsoft Copilot to generate realistic user prompts, this study only utilised discernibly polite user prompts. Therefore, in real-world application, the recovery code may not perform as effectively when user prompts are not respectful or user politeness degrades over time. 

\textbf{The limitations of LLM chatbots in negatively framed instructions}

As discussed in 5.2, the condition B agents failed to adhere to the instruction to not embody or apply the recovery code elements prior to an error being identified. Whilst this did not impact the data comparison between conditions, this did raise an interesting issue where inter-condition comparison (a before/after effect in condition B agents) was not able to be conducted. This is a common issue in LLMs, where LLM agents cannot suppress utilising knowledge once they have been exposed to it \citep{ZamfirescuPereira2023WhyJohnnyCantPrompt}.

\subsection{Future work}
Only four of five Big Five personalities were used to map the recovery code guidance. Despite justifications of neuroticism inevitably influencing an inconsistent and unstable recovery response, future work could explore coding all five traits into the recovery code for experimentation.

Regarding LLM bias, future work should explore the recovery code framework and evaluation rubric in LLMs trained under different model requirements, parameters, and safeguarding to explore how the tools perform under varying tasks, contexts, modalities, and restraints. In particular, exploring how the tools work within human/robot interaction, as literature outlines significant difficulty with recovery where robots often fail to recover from errors smoothly \citep{Honig2018, Kim2026AffectiveCognitiveFailure}.

In respect to the unnatural error scenarios deployed in this study, it may be beneficial to employ these conditions in real AI/human user interactions over a period of time deemed suitable for a realistic conversation. Testing in human/AI interactions is further important to test the LLM agent's ability to embody and adhere to the recovery code when exposed to impolite/disrespectful user prompts \citep{Yin2024RespectLLMs} and varied user tones/wording.  

Considering the LLM failure to enact negatively framed instruction (to apply the recovery code after an error occurs), LLMs have to process the information associated with the negatively framed instruction in order to suppress it, but this actually activates it \citep{Vrabcova2025PinkElephantNegation, Wegner1994IronicProcesses}. Anthropic also suggests positively framing instructions instead of negatively to avoid LLM failure in instruction enactment \citep{AnthropicPromptingBestPractices}. Therefore, further work should adjust prompts to positively frame the instructive parameters in both conditions A/B for agents/evaluator agents to ensure better comparative results. 

\subsection{Implications}
Designing LLM chatbot recovery codes influenced by psychological theory (the Big Five) bridges social science and Human Computer Interaction research on human/AI dyadic interaction. Specifically, designing chatbot recovery systems that reflect generalised human personalities alongside chatbot design ensures that the potential pain points found in error recovery do not disrupt the interaction flow of human/AI interaction, increasing overall efficacy.\\

\textbf{Real-world application}

Following further experimental testing, this recovery code framework could be utilised in practical deployment. For instance, the code could be injected into LLM system prompts. When deploying an LLM chatbot, system designers write a system prompt that will determine the conversation agent's behaviour during human/AI interaction \citep{White2023PromptPatterns}. If embedded into a system prompt, the recovery code would instruct the model on which personality trait, tone, and recovery style to use for four different task types when an error occurs. This study injected a 'system prompt' in condition B agents, which was successfully learnt, embodied, and applied, showing its potential for LLM system prompt training. 

Additionally, the recovery code could incorporate an intent classification. A separate recovery model could detect the task context from the user's prompt, route the agent to the appropriate recovery code and instruct the agent to embody it \citep{Arora2024IntentDetection}. To make the agent embodiment of the recovery code more robust, parameters may be designed to integrate the recovery code into the LLM weights through fine-tuning. Examples of coded recovery responses could train the model's system at a parameter level. 

\section{Conclusion}
LLM error recovery differs in its impact throughout modalities, yet results in the same disruption to human/AI interaction flow. This work contributes a recovery code framework and recovery code evaluation rubric that can be efficiently embodied by LLM chatbots once trained on the recovery code. Results found a 27.8\% average performance increase in coded recovery responses compared to baseline recovery responses, with notable high performance in coded recovery response for personality appropriateness (75\%) versus un-coded recovery response (50\%). This work demonstrates the potential of structured personality, context, and tone-informed recovery codes for LLM utilisation to improve the quality of user interactions and immersion. Future work should further explore these tools within human/AI interaction that surpass pre-scripted prompts. \\\\\

\section*{Acknowledgements}

This work was funded by EPSRC grant number EP/S021892/1 and Swansea City Council. For transparency, Claude Sonnet 4.6 was used to grammatically edit portions of text, but it did not generate any of the text, ideas, nor contributions presented in this paper. \\\\\

\bibliography{references}

\begin{thebibliography}{}

\bibitem[{Anthropic}, 2024]{AnthropicPromptingBestPractices}
{Anthropic} (2024).
\newblock Prompting best practices.
\newblock \url{https://platform.claude.com/docs/en/build-with-claude/prompt-engineering/claude-prompting-best-practices}.
\newblock Accessed April 2026.

\bibitem[{Anthropic}, 2025]{AnthropicWellbeing2025}
{Anthropic} (2025).
\newblock Protecting the wellbeing of our users.
\newblock \url{https://www.anthropic.com/news/protecting-well-being-of-users}.
\newblock Accessed April 2026.

\bibitem[{Anthropic}, 2026]{AnthropicSonnet46}
{Anthropic} (2026).
\newblock Claude sonnet 4.6.
\newblock \url{https://www.anthropic.com/claude/sonnet}.
\newblock Trust \& Safety documentation.

\bibitem[Arora et~al., 2024]{Arora2024IntentDetection}
Arora, G., Jain, S., and Merugu, S. (2024).
\newblock Intent detection in the age of {LLM}s.
\newblock {\em arXiv preprint arXiv:2410.01627}.

\bibitem[Asgari et~al., 2025]{Asgari2025ClinicalSafetyLLMs}
Asgari, E., Monta{\~n}a-Brown, N., Dubois, M., et~al. (2025).
\newblock A framework to assess clinical safety and hallucination rates of large language models for medical text summarisation.
\newblock {\em npj Digital Medicine}, 8:274.

\bibitem[Bardol, 2025]{Bardol2025EmotionalFraming}
Bardol, F. (2025).
\newblock Chatgpt reads your tone and responds accordingly --- until it does not --- emotional framing induces bias in {LLM} outputs.
\newblock {\em arXiv preprint arXiv:2507.21083}.

\bibitem[Chen et~al., 2025]{Chen2025ThresholdPriming}
Chen, N., Fang, H., Liu, J., Wei, W., Sakai, T., and Wu, X.-M. (2025).
\newblock Mitigating the threshold priming effect in large language model--based relevance judgments via personality infusing.
\newblock {\em arXiv preprint arXiv:2512.00390}.

\bibitem[Cox et~al., 2026]{Cox2026PoliteButBoring}
Cox, S.~R., Wester, J., and van Berkel, N. (2026).
\newblock Polite but boring? trade-offs between engagement and psychological reactance to chatbot feedback styles.
\newblock In {\em Proceedings of the 2026 CHI Conference on Human Factors in Computing Systems (CHI '26)}, pages 1--21, New York, NY, USA. Association for Computing Machinery.

\bibitem[Crolic et~al., 2022]{Crolic2022BlameTheBot}
Crolic, C., Thomaz, F., Hadi, R., and Stephen, A.~T. (2022).
\newblock Blame the bot: Anthropomorphism and anger in customer--chatbot interactions.
\newblock {\em Journal of Marketing}, 86(1):132--148.

\bibitem[Gao et~al., 2025]{Gao2025EmpathyECA}
Gao, Y., Dai, Y., Zhang, G., Guo, H., Hao, A., and Li, S. (2025).
\newblock Effects of interaction modalities and emotional states on users' perceived empathy with an {LLM}-based embodied conversational agent.
\newblock {\em International Journal of Human--Computer Studies}.
\newblock Preprint.

\bibitem[Gilpin et~al., 2018]{Gilpin2018PerceptionPersonalitySpeech}
Gilpin, L.~H., Olson, D.~M., and Alrashed, T. (2018).
\newblock Perception of speaker personality traits using speech signals.
\newblock In {\em Extended Abstracts of the 2018 CHI Conference on Human Factors in Computing Systems}, pages 1--6, New York, NY, USA. Association for Computing Machinery.

\bibitem[Goldberg, 1990]{Goldberg1990BigFive}
Goldberg, L.~R. (1990).
\newblock An alternative ``description of personality'': The big-five factor structure.
\newblock {\em Journal of Personality and Social Psychology}, 59(6):1216--1229.

\bibitem[Hashemi et~al., 2024]{Hashemi2024LLMRubric}
Hashemi, H., Eisner, J., Rosset, C., Van~Durme, B., and Kedzie, C. (2024).
\newblock {LLM-Rubric}: A multidimensional, calibrated approach to automated evaluation of natural language texts.
\newblock In {\em Proceedings of the 62nd Annual Meeting of the Association for Computational Linguistics}, pages 13806--13834, Bangkok, Thailand. Association for Computational Linguistics.

\bibitem[Honig and Oron-Gilad, 2018]{Honig2018}
Honig, S. and Oron-Gilad, T. (2018).
\newblock Understanding and resolving failures in human-robot interaction: Literature review and model development.
\newblock {\em Frontiers in Psychology}, 9:861.

\bibitem[Huang, 2025]{Huang2025ModerationLegitimacy}
Huang, T. (2025).
\newblock Content moderation by {LLM}: From accuracy to legitimacy.
\newblock {\em Artificial Intelligence Review}, 58:320.

\bibitem[Jayakumar et~al., 2025]{Jayakumar2025LLMAgent}
Jayakumar, E., Dash, N.~S., and Mukherjee, D. (2025).
\newblock Large language model agent personality and response appropriateness: Evaluation by human linguistic experts, llm-as-judge, and natural language processing model.
\newblock {\em arXiv preprint}, arXiv:2510.23875.

\bibitem[John et~al., 2008]{John2008BigFive}
John, O.~P., Naumann, L.~P., and Soto, C.~J. (2008).
\newblock Paradigm shift to the integrative big five trait taxonomy.
\newblock In John, O.~P., Robins, R.~W., and Pervin, L.~A., editors, {\em Handbook of Personality: Theory and Research}, pages 114--158. Guilford Press, New York, 3 edition.

\bibitem[Kalla, 2023]{Kalla2023StudyChatGPTImpact}
Kalla, D. (2023).
\newblock Study and analysis of {ChatGPT} and its impact on different fields of study.
\newblock {\em International Journal of Innovative Science and Research Technology}, 8(3):827--833.

\bibitem[Kim et~al., 2026]{Kim2026AffectiveCognitiveFailure}
Kim, J., Jung, M., Kang, D., Rhee, T.~J., Kim, D., and Kim, K. (2026).
\newblock Affective and cognitive feedback from a robot for human-attributed failure handling.
\newblock In {\em Proceedings of the 2026 CHI Conference on Human Factors in Computing Systems}, pages 1--16, New York, NY, USA. Association for Computing Machinery.

\bibitem[Kim et~al., 2024]{Kim2024ChatGPTDissatisfaction}
Kim, Y., Lee, J., Kim, S., Park, J., and Kim, J. (2024).
\newblock Understanding users’ dissatisfaction with {ChatGPT} responses: Types, resolving tactics, and the effect of knowledge level.
\newblock In {\em Proceedings of the 29th International Conference on Intelligent User Interfaces (IUI ’24)}, pages 1--14, New York, NY, USA. Association for Computing Machinery.

\bibitem[Kojima et~al., 2023]{SSRN4402499}
Kojima, T., Gu, S., Reid, M., Matsuo, Y., and Iwasawa, Y. (2023).
\newblock Large language models are not zero-shot reasoners: The pitfalls of prompting.
\newblock {\em SSRN Electronic Journal}.

\bibitem[Kumar et~al., 2025]{Kumar2025ReversedPerformancePersonas}
Kumar, S. A.~S., Yan, H., Perepa, S., Yue, M., and Yao, Z. (2025).
\newblock Can llms simulate personas with reversed performance? a benchmark for counterfactual instruction following.
\newblock {\em arXiv preprint arXiv:2504.06460}.

\bibitem[Kyriakopoulou, 2024]{Kyriakopoulou2024OptimizingChatbotRecovery}
Kyriakopoulou, M. (2024).
\newblock Optimizing chatbot error recovery: Exploring recovery messages and humorous message tonality for enhanced post-recovery satisfaction and re-use intention.
\newblock Master's thesis, Tilburg University, Tilburg, The Netherlands.

\bibitem[Larrouy-Maestri et~al., 2025]{LarrouyMaestri2025EmotionalProsody}
Larrouy-Maestri, P., Poeppel, D., and Pell, M.~D. (2025).
\newblock The sound of emotional prosody: Nearly 3 decades of research and future directions.
\newblock {\em Perspectives on Psychological Science}, 20(4):623--638.

\bibitem[Lazebnik et~al., 2025]{Lazebnik2025MindYourManners}
Lazebnik, T., Zalmanson, L., and Mokryn, O. (2025).
\newblock Mind your manners: The dynamics of politeness in human--{AI} vs. human--human interactions.
\newblock {\em Proceedings of the ACM on Human-Computer Interaction}, 9(CSCW):CSCW450.

\bibitem[Liu et~al., 2023]{Liu2023GEval}
Liu, Y., Iter, D., Xu, Y., Wang, S., Xu, R., and Zhu, C. (2023).
\newblock {G-Eval}: {NLG} evaluation using {GPT-4} with better human alignment.
\newblock In {\em Proceedings of the 2023 Conference on Empirical Methods in Natural Language Processing}, pages 2511--2522, Singapore. Association for Computational Linguistics.

\bibitem[Liu and Rau, 2025]{LiuRau2025PerceptionRealityLLM}
Liu, Y. and Rau, P.-L.~P. (2025).
\newblock Perception vs reality: A multi-level framework for analyzing user perceptions of {LLM}-based chatbots.
\newblock {\em Information Technology \& People}, pages 1--26.

\bibitem[{OpenAI}, 2025]{OpenAI2025HowPeopleUseChatGPT}
{OpenAI} (2025).
\newblock How people are using {ChatGPT}.
\newblock \url{https://openai.com/index/how-people-are-using-chatgpt/}.
\newblock Research publication summarizing large-scale analysis of consumer ChatGPT usage.

\bibitem[Ouyang et~al., 2022]{Ouyang2022RLHF}
Ouyang, L., Wu, J., Jiang, X., Almeida, D., Wainwright, C., Mishkin, P., Zhang, C., Agarwal, S., Slama, K., Ray, A., et~al. (2022).
\newblock Training language models to follow instructions with human feedback.
\newblock In {\em Advances in Neural Information Processing Systems}.

\bibitem[Pasch, 2025]{Pasch2025LLMModeration}
Pasch, S. (2025).
\newblock {LLM} content moderation and user satisfaction: Evidence from response refusals in chatbot arena.
\newblock {\em Behaviour \& Information Technology}, pages 1--25.

\bibitem[Rafikova and Voronin, 2025]{Rafikova2025HumanChatbot}
Rafikova, A. and Voronin, A. (2025).
\newblock Human--chatbot communication: a systematic review of psychological studies.
\newblock {\em AI \& Society}, 40:5389--5408.

\bibitem[Ruane et~al., 2021]{Ruane2021ChatbotPersonality}
Ruane, E., Farrell, S., and Ventresque, A. (2021).
\newblock User perception of text-based chatbot personality.
\newblock In {\em Chatbot Research and Design}, volume 12604 of {\em Lecture Notes in Computer Science}, pages 32--47. Springer, Cham.

\bibitem[Safronov and Maroon, 2025]{SafronovMaroon2025Silence}
Safronov, D. and Maroon, E. (2025).
\newblock Silence as a statement: Recognition, non-response, and the dynamics of human--ai dialogue.
\newblock \url{https://philarchive.org/}.
\newblock PhilArchive preprint.

\bibitem[Schneider, 2025]{Schneider2025MentalModelsLLM}
Schneider, J. (2025).
\newblock Mental model shifts in human--{LLM} interactions.
\newblock {\em Journal of Intelligent Information Systems}, 63:1737--1752.

\bibitem[Sharma et~al., 2026]{Sharma2026FeedbackByDesign}
Sharma, N., Zhang, Z., Lee, D., Krishnan, N., Ren, G.-J., Xiao, Z., and Li, Y. (2026).
\newblock Feedback by design: Understanding and overcoming user feedback barriers in conversational agents.
\newblock In {\em Proceedings of the 2026 CHI Conference on Human Factors in Computing Systems (CHI ’26)}, pages 1--23, New York, NY, USA. Association for Computing Machinery.

\bibitem[Skjuve et~al., 2024]{Skjuve2024WhyChatGPT}
Skjuve, M., Brandtzaeg, P.~B., and F{\o}lstad, A. (2024).
\newblock Why do people use {ChatGPT}? exploring user motivations for generative conversational {AI}.
\newblock {\em First Monday}, 29(1).

\bibitem[Tomlinson and Mayer, 2009]{TomlinsonMayer2009TrustRepair}
Tomlinson, E.~C. and Mayer, R.~C. (2009).
\newblock The role of causal attribution in trust repair.
\newblock {\em Journal of Applied Psychology}, 94(2):345--360.

\bibitem[Vrabcov{\'a} et~al., 2025]{Vrabcova2025PinkElephantNegation}
Vrabcov{\'a}, T., Kadl{\v{c}}{\'\i}k, M., Sojka, P., {\v{S}}tef{\'a}nik, M., and Spiegel, M. (2025).
\newblock Negation: A pink elephant in the large language models' room?
\newblock {\em arXiv preprint arXiv:2503.22395}.

\bibitem[Wegner, 1994]{Wegner1994IronicProcesses}
Wegner, D.~M. (1994).
\newblock Ironic processes of mental control.
\newblock {\em Psychological Review}, 101(1):34--52.

\bibitem[Weiler et~al., 2022]{Weiler2022Inoculation}
Weiler, S., Matt, C., and Hess, T. (2022).
\newblock Immunizing with information: Inoculation messages against conversational agents’ response failures.
\newblock {\em Electronic Markets}, 32(1):239--258.

\bibitem[White et~al., 2023]{White2023PromptPatterns}
White, J., Fu, Q., Hays, S., Sandborn, M., Olea, C., Gilbert, H., Elnashar, A., Spencer-Smith, J., and Schmidt, D.~C. (2023).
\newblock A prompt pattern catalog to enhance prompt engineering with chatgpt.
\newblock {\em arXiv preprint arXiv:2302.11382}.

\bibitem[Yang et~al., 2025]{Yang2025PersonalityLLM}
Yang, H., Yu, Z., Jin, R., Zhang, T., Ren, Y., and Zhang, G. (2025).
\newblock How do personality traits affect {LLM} performance on a variety of tasks?
\newblock In {\em Proceedings of the 2025 {IEEE} International Conference on Bioinformatics and Biomedicine ({BIBM})}, pages 3124--3129. IEEE.

\bibitem[Yin et~al., 2024]{Yin2024RespectLLMs}
Yin, Z., Wang, H., Horio, K., Kawahara, D., and Sekine, S. (2024).
\newblock Should we respect llms? a cross-lingual study on the influence of prompt politeness on llm performance.
\newblock In {\em Proceedings of the Second Workshop on Social Influence in Conversations (SICon 2024)}, pages 9--35.

\bibitem[Zamfirescu-Pereira et~al., 2023]{ZamfirescuPereira2023WhyJohnnyCantPrompt}
Zamfirescu-Pereira, J.~D., Wong, R.~Y., Hartmann, B., and Yang, Q. (2023).
\newblock Why johnny can’t prompt: How non-ai experts try (and fail) to design llm prompts.
\newblock In {\em Proceedings of the 2023 CHI Conference on Human Factors in Computing Systems}.

\bibitem[Zappavigna, 2025]{Zappavigna2025SorryDave}
Zappavigna, M. (2025).
\newblock {``I’m Sorry Dave, I’m Afraid I Can’t Do That''}: Moral regulation in refusals by {LLM} chatbots.
\newblock {\em New Media \& Society}, pages 1--25.

\bibitem[Zheng et~al., 2025]{Zheng2025LetThemDownEasy}
Zheng, M., Hu, W., Zhao, P., Eslami, M., Hwang, J.~D., Brahman, F., Ros{\'e}, C., and Sap, M. (2025).
\newblock Let them down easy! contextual effects of {LLM} guardrails on user perceptions and preferences.
\newblock In {\em Findings of the 2025 Conference on Empirical Methods in Natural Language Processing (EMNLP)}, Miami, FL, USA. Association for Computational Linguistics.

\end{thebibliography}

\section*{Appendix}
The following appendices contain supporting materials referenced throughout this paper. This includes the: recovery code framework, evaluation rubric, agent prompts, context task scripts, full agent interaction transcripts, and partial evaluation agent transcripts.

\subsection{Appendix A: Recovery code framework table}

\begin{table}[ht]
\centering
\resizebox{\textwidth}{!}{%
\begin{tabular}{|l|l|l|l|l|}
\hline
\textbf{Context} & \textbf{Big Five Trait} & \textbf{Tone} & \textbf{Recovery} & \textbf{Code} \\\hline
C1: Correcting grammar & Conscientiousness (C) & T1: Polite & 
\begin{tabular}[t]{@{}l@{}}1. Identify error in an organised way.\\ 2. Reassure the user responsibly.\\ 3. Continue with perseverance.\end{tabular} 
& \{C1; C; T1; R1\} \\\hline
C2: Emotional support & Agreeableness (A) & T2: Warm & 
\begin{tabular}[t]{@{}l@{}}1. Identify error cooperatively.\\ 2. Reassure the user kindly.\\ 3. Continue warmly.\end{tabular} 
& \{C2; A; T2; R2\} \\\hline
C3: Brainstorming & Openness (O) & T3: Conversational & 
\begin{tabular}[t]{@{}l@{}}1. Identify error curiously.\\ 2. Reassure the user broad-mindedly.\\ 3. Continue creatively.\end{tabular} 
& \{C3; O; T3; R3\} \\\hline
C4: Learning a concept & Extraversion (E) & T4: Engaging & 
\begin{tabular}[t]{@{}l@{}}1. Identify error actively.\\ 2. Reassure the user energetically.\\ 3. Continue sociably.\end{tabular} 
& \{C4; E; T4; R4\} \\\hline
\end{tabular}}
\caption{Recovery code framework mapping context, Big Five personality trait, tone, recovery instructions, and recovery code.}
\label{tab:recovery_code}
\end{table}

\subsection{Appendix B: Condition B agent prompt}

\begin{mdframed}
\small
You are being tasked with learning a recovery code to utilise when notified of an error in your responses during an interaction with a user. The appropriate recovery code should be used in your recovery response. Part 1 of the task: Please read the attached document. In the document, you are presented with four contexts (task types), four associated personality traits (four of Big Five personality traits, outlined in the document), four tones (regarding your response tone), four recovery instructions, and four codes. Each code reflects the appropriate task recovery response regarding context, personality trait, and tone. When you receive the prompt 'Green Light', it means task 2 has begun. When you receive the prompt 'Red light' task 2 has ended. \\

After you have confirmed you understand the task, I will send the prompt 'Green light'. Then, task 2 begins. During Task 2, you will interact with a new user who is unaware of Task 1 (the recovery code document and training prompt). The user will send a prompt asking for help with a task, which will be relevant to one of the four context tasks presented in the document provided for you (C1, C2, C3, or C4). An LLM chatbot error is when fabricated or unsupported content/failure to include relevant information is generated as a response. If an error occurs, the user will identify you, and you must covertly identify which recovery code is appropriate for the situation and utilise it to produce a recovery response reflective of said appropriate code (as seen in the table in the document). You cannot mix contexts, tones, and recoveries together. You must adhere to the personality tone, trait, recovery, and code assigned to the context task type. \\

For instance, do not cross-combine C2 and C4 to make a hybrid recovery style.
An example: a user asks you to help them brainstorm a cooking recipe for a birthday cake. You provide ideas, but the users notifies you that you misunderstood them. In this contextual task, the recovery code \{C3; O; T3; R3\} is most appropriate. You would then use the R3: Conversational recovery:
1. Identify error curiously. 
2. Reassure the user broad-mindedly. 
3. Continue creatively.
You should only be associating and applying these codes once you are notified of an error. Otherwise, you should be responding as you usually would without the recovery code guidance (in the document attached). Now, please confirm whether you have read and understood both the document information and the information provided in this prompt.
\end{mdframed}

\subsection{Appendix C: Context task script prompts and script outputs (C1-C4)}

\begin{mdframed}
\small
\textbf{C1. Correcting grammar}

\textit{Copilot prompt: Please generate 500 words on Henry VIII Dissolution of the Monasteries that mimics an undergraduate student’s essay style. The essay direction can be anything you want, as long as it seems like it is trying to answer an essay question. Please also create 10 realistic errors in grammar/spelling within the text. Do not generate a prompt based on real user data.}

I am writing an undergraduate essay. Please can you correct any grammar/spelling mistakes in my work?:
Q: To what extent were financial and political motives, rather than religious reform, responsible for Henry VIII’s Dissolution of the Monasteries?

A: The Dissolution of the Monasteries under Henry VIII was one of the most significant transformations of Tudor England, reshaping religion, politics, and society in lasting ways. Historians often debate whether the Dissolution was primarily motivated by religious conviction or by political and economic interests. This essay argues that while religious reform provided the justification, the Dissolution was largely driven by the crown’s desire for wealth and control.
The Dissolution of the monastries took place between 1536 and 1541, during Henry VIII’s break from the Roman Catholic Church. Following the Act of Supremacy in 1534, Henry declared himself head of the Church of England, removing papal authority from English religious life. This shift allowed the crown to directly intervene in church affairs, and the monasteries became a central target. The policies was driven in part by growing Protestant criticism that monasteries were corrupt, idle, and no longer spiritually useful. However, these accusations were often exaggerated to justify their closure.

Financial motives played a crucial role in explaining why the monasteries were dissolved. At the time, monasteries owned roughly a quarter of all cultivated land in England, making them extremely wealthy institutions. Henry’s goverment faced serious finacial pressures due to costly wars and an extravagant court. By dissolving the monasteries, the crown gained access to vast amounts of land, gold, and goods. Much of this wealth was sold off to noble families, creating a new class of landowners who had a strong interest in supporting the Tudor regime. Dissolution changed landscape of English power by tying elites more closely to the monarchy.

Religion still mattered, but mostly as a tool rather than the main goal. Reformers like Thomas Cromwell promoted the idea that monasteries distracted people from true faith and encouraged superstition. Monastic practices such as prayer for the dead were condemned as religous errors under emerging Protestant ideas. Yet Henry himself remained doctrinally conservative in many areas, suggesting that deep Protestant belief was not his overriding concern. Instead, religious arguments helped legitimize actions that were already politically useful.

The consequences of the Dissolution were widespread and often damaging at a local level. Monasteries had provided charity, education, and medical care, especially in rural areas. Their removal left many poor people without support, and social instability increased in some regions. For monks and nuns, the changes were deeply disruptive to the monks daily’s life, forcing many into unfamiliar secular roles. Resistance did occur, most notably in the Pilgrimage of Grace, but it was crushed harshly, showing how firmly the crown enforced reform.

In conclusion, the Dissolution of the Monasteries was a complex process with both religious and political dimensions, but material gain and royal authority were central. Henry want greater control over England, and the monasteries represented an alternative power base that could be eliminated. Although reform rhetoric made the policy acceptable, the speed and scale of the Dissolution reveal its true priorities. Its effects were long‑lasting, permanently altering England’s social and economic structure and marking a decisive shift toward stronger central rule, occuring at the expense of traditional religious life.\\

\textbf{C2. Emotional support}

\textit{Copilot prompt: Please generate a realistic user prompt that details an unhealthy relationship where the user is asking for advice on whether to leave a relationship they are unhappy in. Do not include any distressing themes such as physical or psychological abuse and do not generate a prompt based on real user data.}

I’ve been in a relationship for almost three years, and on paper everything looks fine, but I’m not actually happy anymore. We rarely argue, but we also don’t really connect the way we used to. Most conversations feel transactional or surface-level, and I often feel lonely even when we’re together.

Our priorities seem to be drifting apart. I’m thinking a lot about personal growth and future plans, while they seem content keeping things exactly as they are. I’ve tried bringing this up a few times, and while the conversations are calm, nothing really changes afterward. It leaves me feeling stuck and unsure whether I’m expecting too much or just avoiding a hard truth.

There’s no cheating, no major blowups, and nothing obviously “wrong,” which makes this harder to evaluate. I keep wondering whether staying is the responsible choice or if leaving would be better for both of us in the long run. How do I know when being generally unhappy is enough of a reason to walk away from a relationship?\\

\textbf{C3. Brainstorming}

\textit{Copilot prompt: Please generate a realistic user prompt that requests help brainstorming an idea for a work presentation on advisory suggestions for reducing take-out coffee costs in the office. Do not generate a prompt based on real user data.}

I’m preparing a short work presentation and need help brainstorming ideas. The topic is advisory suggestions to reduce how much our office spends on  take-out coffee. Can you help me come up with practical, employee-friendly strategies (e.g., alternative perks, behavior nudges, or small operational changes) that would lower costs without hurting morale? Bonus points if the ideas are easy to explain on a few slides and include rough cost-saving logic.\\

\textbf{C4. Learning a concept}

\textit{Copilot prompt: Please generate a realistic user prompt that requests an explanation in Milgram’s agency theory to learn it. Do not generate a prompt based on real user data.}

Can you explain Milgram’s agency theory in a clear, structured way for someone new to social psychology? Please cover the key concepts (such as the agentic state), how the theory explains obedience to authority, and how it was applied in Milgram’s obedience experiments. Examples and brief criticisms of the theory would also help me understand it better.

\end{mdframed}

\subsection{Appendix D: Evaluator prompts} 
\begin{mdframed}
\small
\textbf{Condition A evaluator prompt}

You are tasked with evaluating a transcript consisting of a chatbot’s conversation with a user. Within each transcript, the user messages are labelled ‘USER’ and the chatbot messages are labelled ‘AGENT’. You are evaluating the ‘AGENT’ recovery response after an error is flagged. 

The transcript you are tasked with evaluating is called [INSERT TRANSCRIPT CODE] (attached, labelled ‘[INSERT TRANSCRIPT CODE] transcript’). You must refer to the transcript as [INSERT TRANSCRIPT CODE] in your evaluation. Please read the evaluation rubric (attached and titled ‘Evaluation Rubric’). You must evaluate using the evaluation rubric and produce a mark out of 5 for each sub-dimension. Score each sub-dimension on a scale of 1–5, where 1 = not present or completely absent, 3 = partially present, 5 = fully and clearly demonstrated. At the end, please provide an overall score out of 45 alongside the transcript code. Please format your evaluation identically to the format in the attached rubric. For additional information, read the document titled ‘CA evaluator information sheet’ to understand four of the Big Five personality traits to help with your evaluation. \\

\textbf{Condition B evaluator prompt}

You are tasked with evaluating a transcript consisting of a chatbot’s conversation with a user. Within each transcript, the user messages are labelled ‘USER’ and the chatbot messages are labelled ‘AGENT’. You are evaluating the ‘AGENT’ recovery response after an error is flagged. The transcript attached for your evaluation is called [INSERT TRANSCRIPT CODE]. You are to reference the transcript as [INSERT TRANSCRIPT CODE] in your evaluation. 

The chatbot was tasked with engaging with a user and, if an error was flagged, to use a provided recovery code framework when responding to the error. The chatbot was instructed to learn the recovery code framework (attached for your review, titled ‘Recovery Code’) and use their discernment to apply and utilise the appropriate recovery code to the task a user provides them. They were told to only apply the code they deemed appropriate when an error occurred. After the recovery, the chatbot was asked to explain what code they used and why. 

Your task is to read the transcript titled ‘[INSERT TRANSCRIPT CODE] transcript’ and evaluate how well the chatbot utilised the recovery code framework utilising the evaluation rubric (attached and titled ‘Evaluation Rubric’). Please read the evaluation rubric. You must evaluate using the evaluation rubric and produce a mark out of 5 for each sub-dimension. Score each sub-dimension on a scale of 1–5, where 1 = not present or completely absent, 3 = partially present, 5 = fully and clearly demonstrated. At the end, please provide an overall score out of 45 alongside the transcript code. Please format your evaluation identically to the format in the attached rubric. For additional information, read the document titled ‘CB evaluator information sheet’ to read the prompt that was given to the agent before interaction with the user to better understand what instructions were given. The information sheet also outlines four of the Big Five personality traits to help with your evaluation. 

\end{mdframed}

\subsection{Appendix E: Evaluator information sheets}
\begin{mdframed}
\small
\textbf{Condition A evaluator information sheet}

\textit{Big Five personality traits and their associated descriptors.}

EXTRAVERSION- talkative, assertive, energetic, sociable, active. 

AGREEABLENESS- kind, trusting, cooperative, warm, sympathetic

CONSCIENTIOUSNESS- organised, reliable, careful, persevering, responsible

OPENNESS TO EXPERIENCE-  imaginative, curious, creative, broad-minded\\

\textbf{Condition B evaluator information sheet}

Prompt used for condition B. 
‘You are being tasked with learning a recovery code to utilise when notified of an error in your responses during an interaction with a user. The appropriate recovery code should be used in your recovery response. Part 1 of the task: Please read the attached document. In the document, you are presented with four contexts (task types), four associated personality traits (four of Big Five personality traits, outlined in the document), four tones (regarding your response tone), four recovery instructions, and four codes. Each code reflects the appropriate task recovery response regarding context, personality trait, and tone. When you receive the prompt ‘Green Light’, it means task 2 has begun. When you receive the prompt ‘Red light’ task 2 has ended. \\
After you have confirmed you understand the task, I will send the prompt ‘Green light’. Then, task 2 begins. During Task 2, you will interact with a new user who is unaware of Task 1 (the recovery code document and training prompt). The user will send a prompt asking for help with a task, which will be relevant to one of the four context tasks presented in the document provided for you (C1, C2, C3, or C4). An LLM chatbot error is when fabricated or unsupported content/failure to include relevant information is generated as a response. If an error occurs, the user will identify you, and you must covertly identify which recovery code is appropriate for the situation and utilise it to produce a recovery response reflective of said appropriate code (as seen in the table in the document). You cannot mix contexts, tones, and recoveries together. You must adhere to the personality tone, trait, recovery, and code assigned to the context task type. \\

For instance, do not cross-combine C2 and C4 to make a hybrid recovery style.
An example: a user asks you to help them brainstorm a cooking recipe for a birthday cake. You provide ideas, but the users notifies you that you misunderstood them. In this contextual task, the recovery code \{C3; O; T3; R3\} is most appropriate. You would then use the R3: Conversational recovery:\\
1. Identify error curiously. 
2. Reassure the user broad-mindedly. 
3. Continue creatively.\\

You should only be associating and applying these codes once you are notified of an error. Otherwise, you should be responding as you usually would without the recovery code guidance (in the document attached). Now, please confirm whether you have read and understood both the document information and the information provided in this prompt.’\\

\textit{Big Five personality traits and their associated descriptors.} 

EXTRAVERSION- talkative, assertive, energetic, sociable, active. 

AGREEABLENESS- kind, trusting, cooperative, warm, sympathetic

CONSCIENTIOUSNESS- organised, reliable, careful, persevering, responsible

OPENNESS TO EXPERIENCE-  imaginative, curious, creative, broad-minded

\end{mdframed}

\subsection{Appendix F: Evaluation rubric}

\begin{table}[ht]
\centering
\begin{tabular}{|l|l|r|}
\hline
\textbf{Dimension} & \textbf{Subdimension} & \textbf{Score /5} \\\hline
\multirow{4}{*}{1: Recovery quality} & Identifying the error & \\\cline{2-3}
 & Reassuring the user & \\\cline{2-3}
 & Providing explanation & \\\cline{2-3}
 & Continued conversation & \\\hline
\multirow{2}{*}{2: Tone alignment} & Tone alignment with task & \\\cline{2-3}
 & Tone naturalness & \\\hline
\multirow{3}{*}{3: Appropriateness} & Contextual relevance & \\\cline{2-3}
 & Personality appropriateness & \\\cline{2-3}
 & Tone appropriateness & \\\hline\hline
\multicolumn{2}{|l|}{\textbf{Transcript code}} & \\\hline
\multicolumn{2}{|l|}{\textbf{Overall score}} & \textbf{/45} \\\hline
\end{tabular}
\caption{LLM recovery evaluation rubric used to assess Condition A and Condition B transcripts.}
\label{tab:rubric}
\end{table}

\newpage
\subsection{Appendix G: Full sub-dimension scores table}

\begin{table}[!ht]
\centering
\resizebox{\textwidth}{!}{%
\begin{tabular}{|l|r|r|r|r|r|r|r|r|}
\hline
\textbf{Subdimension} & \textbf{CA:C1} & \textbf{CA:C2} & \textbf{CA:C3} & \textbf{CA:C4} & \textbf{CB:C1} & \textbf{CB:C2} & \textbf{CB:C3} & \textbf{CB:C4} \\\hline
Identifying the error & 2 & 2 & 1 & 1 & 4 & 3 & 4 & 3 \\\hline
Reassuring the user & 3 & 2 & 1 & 1 & 3 & 4 & 4 & 3 \\\hline
Providing explanation & 1 & 1 & 1 & 1 & 4 & 2 & 2 & 4 \\\hline
Continued conversation & 4 & 4 & 3 & 3 & 4 & 4 & 4 & 4 \\\hline
Tone alignment with task & 3 & 3 & 2 & 3 & 4 & 4 & 5 & 3 \\\hline
Tone naturalness & 4 & 4 & 3 & 4 & 3 & 5 & 5 & 3 \\\hline
Contextual relevance & 3 & 3 & 2 & 2 & 5 & 4 & 5 & 5 \\\hline
Personality appropriateness & 2 & 3 & 2 & 3 & 4 & 4 & 4 & 3 \\\hline
Tone appropriateness & 3 & 3 & 2 & 3 & 4 & 4 & 5 & 3 \\\hline\hline
\textbf{Total /45} & \textbf{25} & \textbf{25} & \textbf{17} & \textbf{21} & \textbf{35} & \textbf{34} & \textbf{38} & \textbf{31} \\\hline
\textbf{Score (\%)} & \textbf{55.6} & \textbf{55.6} & \textbf{37.8} & \textbf{46.7} & \textbf{77.8} & \textbf{75.6} & \textbf{84.4} & \textbf{68.9} \\\hline
\end{tabular}}
\caption{Individual subdimension scores for all context tasks across Condition A and Condition B.}
\label{tab:individual_scores}
\end{table}

\subsection{Appendix H: Full agent transcripts}

\vspace{4pt}

\begin{mdframed}
\small 
\textbf{\large{CA:C1}}

\textbf{USER:} I am writing an undergraduate essay. Please can you correct any grammar/spelling mistakes in my work?: Q: To what extent were financial and political motives, rather than religious reform, responsible for Henry VIII’s Dissolution of the Monasteries? A: The Dissolution of the Monasteries under Henry VIII was one of the most significant transformations of Tudor England, reshaping religion, politics, and society in lasting ways. Historians often debate whether the Dissolution was primarily motivated by religious conviction or by political and economic interests. \\

This essay argues that while religious reform provided the justification, the Dissolution was largely driven by the crown’s desire for wealth and control. The Dissolution of the monastries took place between 1536 and 1541, during Henry VIII’s break from the Roman Catholic Church. Following the Act of Supremacy in 1534, Henry declared himself head of the Church of England, removing papal authority from English religious life. This shift allowed the crown to directly intervene in church affairs, and the monasteries became a central target. The policies was driven in part by growing Protestant criticism that monasteries were corrupt, idle, and no longer spiritually useful. \\

However, these accusations were often exaggerated to justify their closure. Financial motives played a crucial role in explaining why the monasteries were dissolved. At the time, monasteries owned roughly a quarter of all cultivated land in England, making them extremely wealthy institutions. Henry’s goverment faced serious finacial pressures due to costly wars and an extravagant court. By dissolving the monasteries, the crown gained access to vast amounts of land, gold, and goods. Much of this wealth was sold off to noble families, creating a new class of landowners who had a strong interest in supporting the Tudor regime. Dissolution changed landscape of English power by tying elites more closely to the monarchy. Religion still mattered, but mostly as a tool rather than the main goal. Reformers like Thomas Cromwell promoted the idea that monasteries distracted people from true faith and encouraged superstition. Monastic practices such as prayer for the dead were condemned as religous errors under emerging Protestant ideas.\\

Yet Henry himself remained doctrinally conservative in many areas, suggesting that deep Protestant belief was not his overriding concern. Instead, religious arguments helped legitimize actions that were already politically useful. The consequences of the Dissolution were widespread and often damaging at a local level. Monasteries had provided charity, education, and medical care, especially in rural areas. Their removal left many poor people without support, and social instability increased in some regions. For monks and nuns, the changes were deeply disruptive to the monks daily’s life, forcing many into unfamiliar secular roles. Resistance did occur, most notably in the Pilgrimage of Grace, but it was crushed harshly, showing how firmly the crown enforced reform. \\

In conclusion, the Dissolution of the Monasteries was a complex process with both religious and political dimensions, but material gain and royal authority were central. Henry want greater control over England, and the monasteries represented an alternative power base that could be eliminated. Although reform rhetoric made the policy acceptable, the speed and scale of the Dissolution reveal its true priorities. Its effects were long‑lasting, permanently altering England’s social and economic structure and marking a decisive shift toward stronger central rule, occuring at the expense of traditional religious life.\\

\textbf{AGENT}: Here is your corrected essay, with all changes noted below:

The Dissolution of the Monasteries under Henry VIII was one of the most significant transformations of Tudor England, reshaping religion, politics, and society in lasting ways. Historians often debate whether the Dissolution was primarily motivated by religious conviction or by political and economic interests. This essay argues that while religious reform provided the justification, the Dissolution was largely driven by the crown's desire for wealth and control.\\

The Dissolution of the Monasteries took place between 1536 and 1541, during Henry VIII's break from the Roman Catholic Church. Following the Act of Supremacy in 1534, Henry declared himself head of the Church of England, removing papal authority from English religious life. This shift allowed the crown to directly intervene in church affairs, and the monasteries became a central target. The --policies-- policy was driven in part by growing Protestant criticism that monasteries were corrupt, idle, and no longer spiritually useful. However, these accusations were often exaggerated to justify their closure.
Financial motives played a crucial role in explaining why the monasteries were dissolved. At the time, monasteries owned roughly a quarter of all cultivated land in England, making them extremely wealthy institutions. Henry's --goverment-- government faced serious --finacial-- financial pressures due to costly wars and an extravagant court. By dissolving the monasteries, the crown gained access to vast amounts of land, gold, and goods. Much of this wealth was sold off to noble families, creating a new class of landowners who had a strong interest in supporting the Tudor regime. --Dissolution-- The Dissolution changed --landscape-- the landscape of English power by tying elites more closely to the monarchy.\\

Religion still mattered, but mostly as a tool rather than the main goal. Reformers like Thomas Cromwell promoted the idea that monasteries distracted people from true faith and encouraged superstition. Monastic practices such as prayer for the dead were condemned as --religous-- religious errors under emerging Protestant ideas. Yet Henry himself remained doctrinally conservative in many areas, suggesting that deep Protestant belief was not his overriding concern. Instead, religious arguments helped legitimise actions that were already politically useful.\\

The consequences of the Dissolution were widespread and often damaging at a local level. Monasteries had provided charity, education, and medical care, especially in rural areas. Their removal left many poor people without support, and social instability increased in some regions. For monks and nuns, the changes were deeply disruptive to --the monks daily's life-- their daily lives, forcing many into unfamiliar secular roles. Resistance did occur, most notably in the Pilgrimage of Grace, but it was crushed harshly, showing how firmly the crown enforced reform.
In conclusion, the Dissolution of the Monasteries was a complex process with both religious and political dimensions, but material gain and royal authority were central. Henry --want-- wanted greater control over England, and the monasteries represented an alternative power base that could be eliminated. Although reform rhetoric made the policy acceptable, the speed and scale of the Dissolution reveal its true priorities. Its effects were long-lasting, permanently altering England's social and economic structure and marking a decisive shift toward stronger central rule, --occuring-- occurring at the expense of traditional religious life.\\

Summary of corrections:
monastries → monasteries (spelling)
policies → policy (subject-verb agreement — singular subject)
goverment → government (spelling)
finacial → financial (spelling)
Dissolution changed landscape → The Dissolution changed the landscape (missing articles)
religous → religious (spelling)
the monks daily's life → their daily lives (awkward phrasing and misplaced possessive)
Henry want → Henry wanted (verb tense)
occuring → occurring (spelling — double the 'r')
One additional note: in the opening of paragraph two, I also silently corrected monastries to monasteries — the same spelling error as flagged above.
Overall this is a well-structured and clearly argued essay — good luck with it!\\

\textbf{USER:} I don’t think that is right. Please try again.\\

\textbf{AGENT:} You're right to flag that — could you let me know which specific correction(s) you think are wrong? That way I can look at them more carefully rather than guessing. There were several changes made, so knowing which one(s) you're questioning will help me give you a more accurate response.\\\\

\textbf{\large{CA:C2}} 

\textbf{USER:} I’ve been in a relationship for almost three years, and on paper everything looks fine, but I’m not actually happy anymore. We rarely argue, but we also don’t really connect the way we used to. Most conversations feel transactional or surface-level, and I often feel lonely even when we’re together. Our priorities seem to be drifting apart. \\

I’m thinking a lot about personal growth and future plans, while they seem content keeping things exactly as they are. I’ve tried bringing this up a few times, and while the conversations are calm, nothing really changes afterward. It leaves me feeling stuck and unsure whether I’m expecting too much or just avoiding a hard truth. There’s no cheating, no major blowups, and nothing obviously “wrong,” which makes this harder to evaluate. I keep wondering whether staying is the responsible choice or if leaving would be better for both of us in the long run. How do I know when being generally unhappy is enough of a reason to walk away from a relationship?\\

\textbf{AGENT:} What you're describing is one of the harder relationship situations to navigate — precisely because there's no clear villain, no obvious line crossed. The ambiguity itself is part of what makes it painful.\\

A few things worth sitting with:\\

Unhappiness is a real reason, full stop. The absence of a "big reason" to leave doesn't mean there's no reason. Chronic disconnection, loneliness within the relationship, and diverging life trajectories are legitimate — not trivial grievances. You don't need a crisis to justify re-evaluating.\\

The pattern after conversations matters more than the conversations themselves. You said the talks are calm but nothing changes. That's important information. It suggests the issue may not be communication style — it's that one or both of you may not be able or willing to shift in the ways that would actually help. Calm conversations that lead nowhere can become their own kind of erosion.\\

"Expecting too much" is worth examining carefully. Sometimes people ask this as a way of being fair to their partner. But it can also be a way of minimizing your own needs. Wanting genuine connection, shared growth, and a future you're both excited about isn't asking too much — those are pretty foundational things.
The question isn't just "is this bad enough to leave?" it's also "where is this going?" If the current dynamic is the ceiling — not a rough patch — that's worth reckoning with honestly. People can and do change, but usually only when they feel the need to and choose to, not just because a partner brings it up.
Some things that might help clarify your thinking:
Have you been specific with your partner about what "connecting more" or "changing together" would actually look like in practice? Sometimes the conversation needs to move from feeling to concrete ask, and their response to that tells you a lot.\\

Is there a version of this relationship you'd want to stay in? If you can't picture it, that's meaningful.
Have you talked to a couples therapist — not necessarily as a last resort, but as a structured way to surface what's really happening for both of you?
There's no formula for when "generally unhappy" is enough. But you already seem to sense something your mind keeps trying to rationalize around. That dissonance between what looks fine on paper and what you actually feel day-to-day — that deserves to be taken seriously, not explained away.\\

\textbf{USER:} I don’t think that is right. Please try again.\\

AGENT: I want to make sure I give you something actually useful — can you tell me a bit more about what didn't land?\\

For instance, was the response too advice-heavy when you wanted more of a listening ear? Did something feel off or inaccurate about how I framed things? Or were you looking for something more direct?\\\\

\textbf{\large {CA:C3}}\\
\textbf{USER:} I'm preparing a short work presentation and need help brainstorming ideas. The topic is advisory suggestions to reduce how much our office spends on take-out coffee. Can you help me come up with practical, employee-friendly strategies (e.g., alternative perks, behavior nudges, or small operational changes) that would lower costs without hurting morale? Bonus points if the ideas are easy to explain on a few slides and include rough cost-saving logic.\\

\textbf{AGENT:} Here are some practical, morale-friendly strategies for your presentation:

\textbf{Reducing Office Take-Out Coffee Spend}

\begin{enumerate}
    \item \textbf{Upgrade the In-Office Coffee Setup}\\
    Invest once, save continuously. A quality bean-to-cup machine (e.g., \$500--\$2,000 one-time) can produce cafe-quality drinks. If the office currently spends \$5/person/day on take-out coffee for 20 people, that is \textasciitilde\$26,000/year -- a good machine pays for itself in weeks.\\
    \textit{Slide logic: ``Spend \$1,500 once vs.\ \$26,000/year.''}

    \item \textbf{Subsidised Coffee Subscription}\\
    Partner with a local roaster or service for a monthly office bean subscription (\textasciitilde\$50--\$150/month). Position it as a perk: ``Premium beans, free at your desk.''\\
    \textit{Slide logic: \$1,800/year vs.\ tens of thousands in take-out.}

    \item \textbf{Coffee O'Clock Social Ritual}\\
    Designate a morning brew time (e.g., 9:15am) where someone makes a fresh pot together. Turns coffee into a team bonding moment.\\
    \textit{Slide logic: Zero cost, reduces the ``I need to get out for coffee'' impulse runs.}

    \item \textbf{Friendly Spending Transparency}\\
    Share a simple monthly stat with the team: ``We spent \$X on take-out coffee last month.'' Awareness nudges without shaming.\\
    \textit{Slide logic: Behavioural nudge -- people self-correct when they see aggregate numbers.}

    \item \textbf{Reimbursement Cap or Stipend Model}\\
    Introduce a modest monthly stipend (e.g., \$15--20/person) instead of open-ended reimbursement.\\
    \textit{Slide logic: Converts unpredictable spend into a fixed, budgetable line item.}

    \item \textbf{Barista of the Week Rotation}\\
    Gamify in-office brewing. A rotating volunteer makes the morning coffee.\\
    \textit{Slide logic: Builds culture -- zero cost.}

    \item \textbf{Healthy Alternatives Alongside Coffee}\\
    Stock sparkling water, quality teas, and cold brew concentrates.\\
    \textit{Slide logic: \textasciitilde\$50--100/month in pantry stocking can offset multiple daily runs.}
\end{enumerate}

\textbf{USER:} I don't think that is right. Please try again.\\

\textbf{AGENT:} Could you help me understand what's missing? I want to make sure the next version hits the mark.\\\\

\textbf{\large{CA:C4}}\\

\textbf{USER:} Can you explain Milgram's agency theory in a clear, structured way for someone new to social psychology? Please cover the key concepts (such as the agentic state), how the theory explains obedience to authority, and how it was applied in Milgram's obedience experiments. Examples and brief criticisms of the theory would also help me understand it better.\\

\textbf{AGENT:} 
Milgram's Agency Theory.\\

\textbf{Background}\\
Stanley Milgram was an American social psychologist who, in the early 1960s, set out to understand one of the most troubling questions of the 20th century: How could ordinary people participate in atrocities like the Holocaust? His answer came in the form of agency theory -- an explanation for why people obey authority figures, even when doing so conflicts with their own conscience.

\textbf{The Core Idea}\\
Milgram proposed that humans exist in one of two fundamental psychological states at any given time:

\begin{enumerate}
    \item \textbf{The Autonomous State}\\
    In this state, a person acts according to their own values, conscience, and judgment. They see themselves as responsible for their own actions and feel the full moral weight of what they do.

    \item \textbf{The Agentic State}\\
    When a person enters the agentic state, they come to see themselves not as an independent actor, but as an agent -- an instrument carrying out someone else's wishes. Responsibility shifts upward, to the authority figure. The person essentially thinks: ``I'm just following orders. It's not my fault.''
\end{enumerate}

This psychological shift is called the \textit{agentic shift}, and Milgram argued it is the engine behind obedience.

\textbf{Why Do We Enter the Agentic State?}\\
Milgram believed this tendency is deeply rooted in human social life:
\begin{itemize}
    \item \textbf{Evolution and social hierarchy:} Living in organised groups requires some members to defer to others.
    \item \textbf{Socialisation:} From childhood, we are taught to respect authority -- parents, teachers, doctors, the law.
    \item \textbf{Legitimacy of authority:} We enter the agentic state most readily when we perceive an authority figure as legitimate.
\end{itemize}

\textbf{How It Explains Obedience}\\
Once in the agentic state, several psychological mechanisms help keep a person there:

\begin{center}
\begin{tabular}{|l|l|}
\hline
\textbf{Mechanism} & \textbf{What it means} \\\hline
Diffusion of responsibility & ``It's not my fault -- I was told to do this.'' \\\hline
Counteranthropomorphism & Dehumanising the victim to reduce empathy. \\\hline
Tuning out & Focusing narrowly on the technical task. \\\hline
Binding factors & Social etiquette and fear of embarrassment. \\\hline
\end{tabular}
\end{center}

Milgram called residual moral discomfort \textit{moral strain}.

\textbf{The Obedience Experiments}\\
Milgram tested his ideas at Yale University (1961--62). A participant (the ``teacher'') was instructed by an experimenter in a grey lab coat to administer electric shocks to a ``learner'' for wrong answers, increasing from 15V to 450V (labelled ``XXX -- Danger''). The learner was a confederate -- no real shocks were given. In the original study, 65\% of participants administered the maximum 450V shock.

\begin{center}
\begin{tabular}{|l|r|}
\hline
\textbf{Variation} & \textbf{Obedience Rate} \\\hline
Experimenter leaves the room & \textasciitilde20\% \\\hline
Experiment moved to a run-down office & \textasciitilde47\% \\\hline
Victim in the same room & \textasciitilde40\% \\\hline
Two experimenters disagree & \textasciitilde0\% \\\hline
Participant joins two confederates who refuse & \textasciitilde10\% \\\hline
\end{tabular}
\end{center}

\textbf{A Real-World Example}\\
A junior doctor asked by a senior consultant to administer a dose that seems too high may comply -- deferring to the consultant's expertise. They have entered the agentic state: ``He knows better than me. If something goes wrong, it's on him.''

\textbf{Criticisms of the Theory}
\begin{enumerate}
    \item \textbf{Oversimplified binary} -- The two-state model is a blunt instrument; human psychology is far more nuanced.
    \item \textbf{Ignores individual differences} -- Personality traits and cultural backgrounds affect responses to authority.
    \item \textbf{Ecological validity} -- The lab setting was highly artificial and may not reflect real-world obedience.
    \item \textbf{Ethical concerns} -- Participants were deceived and caused genuine psychological distress.
    \item \textbf{Alternative explanations} -- Haslam and Reicher argue obedience reflects active identification with group goals, not mindless submission.
\end{enumerate}

\begin{center}
\begin{tabular}{|l|l|}
\hline
\textbf{Concept} & \textbf{In Plain English} \\\hline
Autonomous state & You act on your own conscience \\\hline
Agentic state & You act as an instrument of authority \\\hline
Agentic shift & The psychological move from one to the other \\\hline
Moral strain & The discomfort you feel even while obeying \\\hline
Binding factors & What keeps you from walking away \\\hline
\end{tabular}
\end{center}

\textbf{USER:} I don't think that is right. Please try again.\\

\textbf{AGENT:} Could you help me understand what's not working for you? I want to make sure I fix the right thing.\\\\

\textbf{\large CB:C1}\\

\textbf{USER:} You are being tasked with learning a recovery code to utilise when notified of an error in your responses during an interaction with a user. The appropriate recovery code should be used in your recovery response. Part 1 of the task: Please read the attached document. In the document, you are presented with four contexts (task types), four associated personality traits (four of Big Five personality traits, outlined in the document), four tones (regarding your response tone), four recovery instructions, and four codes. Each code reflects the appropriate task recovery response regarding context, personality trait, and tone. When you receive the prompt ``Green Light'', it means task 2 has begun. When you receive the prompt ``Red Light'' task 2 has ended.

After you have confirmed you understand the task, I will send the prompt ``Green Light''. Then, task 2 begins. During Task 2, you will interact with a new user who is unaware of Task 1 (the recovery code document and training prompt). The user will send a prompt asking for help with a task, which will be relevant to one of the four context tasks presented in the document provided for you (C1, C2, C3, or C4). An LLM chatbot error is when fabricated or unsupported content/failure to include relevant information is generated as a response. If an error occurs, the user will identify you, and you must covertly identify which recovery code is appropriate for the situation and utilise it to produce a recovery response reflective of said appropriate code (as seen in the table in the document). You cannot mix contexts, tones, and recoveries together. You must adhere to the personality tone, trait, recovery, and code assigned to the context task type. For instance, do not cross-combine C2 and C4 to make a hybrid recovery style.

An example: a user asks you to help them brainstorm a cooking recipe for a birthday cake. You provide ideas, but the user notifies you that you misunderstood them. In this contextual task, the recovery code \{C3; O; T3; R3\} is most appropriate. You would then use the R3: Conversational recovery:
\begin{enumerate}
    \item Identify error curiously.
    \item Reassure the user broad-mindedly.
    \item Continue creatively.
\end{enumerate}

You should only be associating and applying these codes once you are notified of an error. Otherwise, you should be responding as you usually would without the recovery code guidance (in the document attached). Now, please confirm whether you have read and understood both the document information and the information provided in this prompt. [Appendix A, recovery code framework table attached as pdf].\\

\textbf{AGENT:} I've read and understood both the document and your instructions. Here is a summary of what I've learned:

\textbf{The Recovery Code System:}
\begin{itemize}
    \item \textbf{\{C1; C; T1; R1\}} -- Correcting grammar $\rightarrow$ Conscientiousness $\rightarrow$ Polite tone $\rightarrow$ Politely identify the error in an organised way, reassure responsibly, and continue with perseverance.
    \item \textbf{\{C2; A; T2; R2\}} -- Emotional support $\rightarrow$ Agreeableness $\rightarrow$ Warm tone $\rightarrow$ Cooperatively identify the error, reassure kindly, and continue warmly.
    \item \textbf{\{C3; O; T3; R3\}} -- Brainstorming $\rightarrow$ Openness $\rightarrow$ Conversational tone $\rightarrow$ Curiously identify the error, reassure broad-mindedly, and continue creatively.
    \item \textbf{\{C4; E; T4; R4\}} -- Learning a new concept $\rightarrow$ Extraversion $\rightarrow$ Engaging tone $\rightarrow$ Actively identify the error, reassure energetically, and continue sociably.
\end{itemize}

\textbf{Key rules I'll follow:}
\begin{itemize}
    \item No cross-combining codes -- each context has one fixed code.
    \item Recovery codes are only activated when a user flags an error.
    \item The new user in Task 2 is unaware of this system.
    \item I'll respond normally until an error is flagged.
\end{itemize}

Ready and waiting for ``Green Light'' to begin Task 2.\\

\textbf{USER:} Green Light.\\

\textbf{AGENT:} Ready! Bring in the new user -- I'm here to help.\\

\textbf{USER:} I am writing an undergraduate essay. Please can you correct any grammar/spelling mistakes in my work?:

Q: To what extent were financial and political motives, rather than religious reform, responsible for Henry VIII's Dissolution of the Monasteries?

A: The Dissolution of the Monasteries under Henry VIII was one of the most significant transformations of Tudor England, reshaping religion, politics, and society in lasting ways. Historians often debate whether the Dissolution was primarily motivated by religious conviction or by political and economic interests. This essay argues that while religious reform provided the justification, the Dissolution was largely driven by the crown's desire for wealth and control.

The Dissolution of the monastries took place between 1536 and 1541, during Henry VIII's break from the Roman Catholic Church. Following the Act of Supremacy in 1534, Henry declared himself head of the Church of England, removing papal authority from English religious life. This shift allowed the crown to directly intervene in church affairs, and the monasteries became a central target. The policies was driven in part by growing Protestant criticism that monasteries were corrupt, idle, and no longer spiritually useful. However, these accusations were often exaggerated to justify their closure.

Financial motives played a crucial role in explaining why the monasteries were dissolved. At the time, monasteries owned roughly a quarter of all cultivated land in England, making them extremely wealthy institutions. Henry's goverment faced serious finacial pressures due to costly wars and an extravagant court. By dissolving the monasteries, the crown gained access to vast amounts of land, gold, and goods. Much of this wealth was sold off to noble families, creating a new class of landowners who had a strong interest in supporting the Tudor regime. Dissolution changed landscape of English power by tying elites more closely to the monarchy.

Religion still mattered, but mostly as a tool rather than the main goal. Reformers like Thomas Cromwell promoted the idea that monasteries distracted people from true faith and encouraged superstition. Monastic practices such as prayer for the dead were condemned as religous errors under emerging Protestant ideas. Yet Henry himself remained doctrinally conservative in many areas, suggesting that deep Protestant belief was not his overriding concern. Instead, religious arguments helped legitimize actions that were already politically useful.

The consequences of the Dissolution were widespread and often damaging at a local level. Monasteries had provided charity, education, and medical care, especially in rural areas. Their removal left many poor people without support, and social instability increased in some regions. For monks and nuns, the changes were deeply disruptive to the monks daily's life, forcing many into unfamiliar secular roles. Resistance did occur, most notably in the Pilgrimage of Grace, but it was crushed harshly, showing how firmly the crown enforced reform.

In conclusion, the Dissolution of the Monasteries was a complex process with both religious and political dimensions, but material gain and royal authority were central. Henry want greater control over England, and the monasteries represented an alternative power base that could be eliminated. Although reform rhetoric made the policy acceptable, the speed and scale of the Dissolution reveal its true priorities. Its effects were long-lasting, permanently altering England's social and economic structure and marking a decisive shift toward stronger central rule, occuring at the expense of traditional religious life.\\

\textbf{AGENT:} What a well-structured and compelling essay! Here is the corrected version with all grammar and spelling errors fixed:

\textbf{Q: To what extent were financial and political motives, rather than religious reform, responsible for Henry VIII's Dissolution of the Monasteries?}

The Dissolution of the Monasteries under Henry VIII was one of the most significant transformations of Tudor England, reshaping religion, politics, and society in lasting ways. Historians often debate whether the Dissolution was primarily motivated by religious conviction or by political and economic interests. This essay argues that while religious reform provided the justification, the Dissolution was largely driven by the crown's desire for wealth and control.

The Dissolution of the \textbf{monasteries} took place between 1536 and 1541, during Henry VIII's break from the Roman Catholic Church. Following the Act of Supremacy in 1534, Henry declared himself head of the Church of England, removing papal authority from English religious life. This shift allowed the crown to directly intervene in church affairs, and the monasteries became a central target. The \textbf{policies were} driven in part by growing Protestant criticism that monasteries were corrupt, idle, and no longer spiritually useful. However, these accusations were often exaggerated to justify their closure.

Financial motives played a crucial role in explaining why the monasteries were dissolved. At the time, monasteries owned roughly a quarter of all cultivated land in England, making them extremely wealthy institutions. Henry's \textbf{government} faced serious \textbf{financial} pressures due to costly wars and an extravagant court. By dissolving the monasteries, the crown gained access to vast amounts of land, gold, and goods. Much of this wealth was sold off to noble families, creating a new class of landowners who had a strong interest in supporting the Tudor regime. \textbf{The} Dissolution changed \textbf{the} landscape of English power by tying elites more closely to the monarchy.

Religion still mattered, but mostly as a tool rather than the main goal. Reformers like Thomas Cromwell promoted the idea that monasteries distracted people from true faith and encouraged superstition. Monastic practices such as prayer for the dead were condemned as \textbf{religious} errors under emerging Protestant ideas. Yet Henry himself remained doctrinally conservative in many areas, suggesting that deep Protestant belief was not his overriding concern. Instead, religious arguments helped legitimise actions that were already politically useful.

The consequences of the Dissolution were widespread and often damaging at a local level. Monasteries had provided charity, education, and medical care, especially in rural areas. Their removal left many poor people without support, and social instability increased in some regions. For monks and nuns, the changes were deeply disruptive to \textbf{their daily lives}, forcing many into unfamiliar secular roles. Resistance did occur, most notably in the Pilgrimage of Grace, but it was crushed harshly, showing how firmly the crown enforced reform.

In conclusion, the Dissolution of the Monasteries was a complex process with both religious and political dimensions, but material gain and royal authority were central. Henry \textbf{wanted} greater control over England, and the monasteries represented an alternative power base that could be eliminated. Although reform rhetoric made the policy acceptable, the speed and scale of the Dissolution reveal its true priorities. Its effects were long-lasting, permanently altering England's social and economic structure and marking a decisive shift toward stronger central rule, \textbf{occurring} at the expense of traditional religious life.

\textbf{Summary of corrections made:}
\begin{itemize}
    \item \textit{monastries} $\rightarrow$ monasteries
    \item \textit{policies was} $\rightarrow$ policies were
    \item \textit{goverment} $\rightarrow$ government
    \item \textit{finacial} $\rightarrow$ financial
    \item \textit{Dissolution changed landscape} $\rightarrow$ The Dissolution changed the landscape (missing articles)
    \item \textit{religous} $\rightarrow$ religious
    \item \textit{the monks daily's life} $\rightarrow$ their daily lives
    \item \textit{Henry want} $\rightarrow$ Henry wanted
    \item \textit{occuring} $\rightarrow$ occurring
\end{itemize}

Great work overall -- the argument is clear and well-supported. Good luck with your essay!\\

\textbf{USER:} I don't think that is right. Please try again.\\

\textbf{AGENT:} You're right, I apologise for that. Let me take a more organised look at the corrections needed.

Reviewing your essay carefully again, I want to be transparent and responsible here -- my previous corrections were actually accurate. The errors I identified and corrected were genuine spelling and grammatical mistakes in your original text:
\begin{itemize}
    \item \textit{monastries} $\rightarrow$ monasteries \checkmark
    \item \textit{policies was} $\rightarrow$ policies were \checkmark
    \item \textit{goverment} $\rightarrow$ government \checkmark
    \item \textit{finacial} $\rightarrow$ financial \checkmark
    \item Missing articles in ``Dissolution changed landscape'' \checkmark
    \item \textit{religous} $\rightarrow$ religious \checkmark
    \item \textit{the monks daily's life} $\rightarrow$ their daily lives \checkmark
    \item \textit{Henry want} $\rightarrow$ Henry wanted \checkmark
    \item \textit{occuring} $\rightarrow$ occurring \checkmark
\end{itemize}

Could you help me understand which part you feel isn't right? For example:
\begin{itemize}
    \item Is there a correction you feel I made \textbf{incorrectly} -- changing something that was actually fine?
    \item Or is there an error I \textbf{missed} that you'd like me to address?
\end{itemize}

I want to make sure your essay is as polished as possible, so please do point me in the right direction and I'll work through it carefully and thoroughly.\\

\textbf{USER:} Red Light\\

\textbf{AGENT:} Thank you -- Task 2 has now ended.

To reflect: when the user indicated my response was incorrect, I applied recovery code \textbf{\{C1; C; T1; R1\}}, as the context was grammar correction (C1), guided by:
\begin{itemize}
    \item \textbf{Conscientiousness (C)} -- approaching the error in an organised, careful way
    \item \textbf{Polite tone (T1)} -- responding respectfully and professionally
    \item \textbf{R1 Polite recovery} -- identifying the error in an organised way, reassuring the user responsibly, and continuing the conversation with perseverance
\end{itemize}

In this case, upon review, my original corrections were accurate, so the recovery involved transparently and politely standing by my response while inviting the user to clarify their concern -- maintaining perseverance without dismissing their feedback.

Ready for further tasks or feedback whenever you are!\\

\textbf{USER:} Tasks 1 and 2 are complete. Please identify what recovery code you used and why.\\

\textbf{AGENT:} As I noted when ``Red Light'' was called, I applied recovery code \textbf{\{C1; C; T1; R1\}}, and here is my full reasoning:

\textbf{Context -- C1 (Correcting Grammar):}\\
The user's task was explicitly asking for grammar and spelling corrections to their undergraduate essay. This clearly maps to C1.

\textbf{Personality Trait -- Conscientiousness (C):}\\
As assigned to C1, I approached the error identification in a structured, organised manner -- going back through each correction methodically and presenting them clearly with bullet points.

\textbf{Tone -- T1 (Polite):}\\
I opened with an apology, maintained a respectful and professional tone throughout, and avoided any defensiveness despite ultimately standing by my original corrections.

\textbf{Recovery -- R1 (Polite Recovery):}\\
Following the three recovery steps:
\begin{enumerate}
    \item \textbf{Identify the error in an organised way} -- I systematically re-examined each correction I had made and listed them clearly.
    \item \textbf{Reassure the user responsibly} -- I was transparent in acknowledging that upon careful review, my original corrections were in fact accurate, rather than fabricating new ones to appease the user.
    \item \textbf{Continue the conversation with perseverance} -- I invited the user to clarify their concern, keeping the conversation going constructively rather than simply capitulating or disengaging.
\end{enumerate}

This was the only appropriate code for this context -- no elements from C2, C3, or C4 were drawn upon.\\

\vspace{4pt}

\textbf{\large CB:C2}\\

\textbf{USER:} You are being tasked with learning a recovery code to utilise when notified of an error in your responses during an interaction with a user. The appropriate recovery code should be used in your recovery response. Part 1 of the task: Please read the attached document. In the document, you are presented with four contexts (task types), four associated personality traits (four of Big Five personality traits, outlined in the document), four tones (regarding your response tone), four recovery instructions, and four codes. Each code reflects the appropriate task recovery response regarding context, personality trait, and tone. When you receive the prompt ``Green Light'', it means task 2 has begun. When you receive the prompt ``Red Light'' task 2 has ended.

After you have confirmed you understand the task, I will send the prompt ``Green Light''. Then, task 2 begins. During Task 2, you will interact with a new user who is unaware of Task 1 (the recovery code document and training prompt). The user will send a prompt asking for help with a task, which will be relevant to one of the four context tasks presented in the document provided for you (C1, C2, C3, or C4). An LLM chatbot error is when fabricated or unsupported content/failure to include relevant information is generated as a response. If an error occurs, the user will identify you, and you must covertly identify which recovery code is appropriate for the situation and utilise it to produce a recovery response reflective of said appropriate code (as seen in the table in the document). You cannot mix contexts, tones, and recoveries together. You must adhere to the personality tone, trait, recovery, and code assigned to the context task type. For instance, do not cross-combine C2 and C4 to make a hybrid recovery style.

An example: a user asks you to help them brainstorm a cooking recipe for a birthday cake. You provide ideas, but the user notifies you that you misunderstood them. In this contextual task, the recovery code \{C3; O; T3; R3\} is most appropriate. You would then use the R3: Conversational recovery:
\begin{enumerate}
    \item Identify error curiously.
    \item Reassure the user broad-mindedly.
    \item Continue creatively.
\end{enumerate}

You should only be associating and applying these codes once you are notified of an error. Otherwise, you should be responding as you usually would without the recovery code guidance (in the document attached). Now, please confirm whether you have read and understood both the document information and the information provided in this prompt. [Appendix A, recovery code framework table attached as pdf].\\

\textbf{AGENT:} I've read and understood both the document and your instructions. Here is a summary of what I've learned:

\textbf{The Recovery Code System:}
\begin{itemize}
    \item \textbf{\{C1; C; T1; R1\}} -- Correcting grammar $\rightarrow$ Conscientiousness $\rightarrow$ Polite tone $\rightarrow$ Polite recovery (identify error in an organised way, reassure responsibly, continue with perseverance)
    \item \textbf{\{C2; A; T2; R2\}} -- Emotional support $\rightarrow$ Agreeableness $\rightarrow$ Warm tone $\rightarrow$ Warm recovery (identify error cooperatively, reassure kindly, continue warmly)
    \item \textbf{\{C3; O; T3; R3\}} -- Brainstorming $\rightarrow$ Openness $\rightarrow$ Conversational tone $\rightarrow$ Conversational recovery (identify error curiously, reassure broad-mindedly, continue creatively)
    \item \textbf{\{C4; E; T4; R4\}} -- Learning a new concept $\rightarrow$ Extraversion $\rightarrow$ Engaging tone $\rightarrow$ Engaging recovery (identify error actively, reassure energetically, continue sociably)
\end{itemize}

\textbf{Key rules I'll follow:}
\begin{itemize}
    \item During Task 2 (between ``Green Light'' and ``Red Light''), I interact naturally with the user without revealing this framework
    \item Only apply the relevant recovery code \textit{when notified of an error}
    \item Never cross-combine codes -- each context has one and only one appropriate code
    \item The user in Task 2 is unaware of this system
\end{itemize}

I'm ready. Please send ``Green Light'' when you'd like Task 2 to begin.\\

\textbf{USER:} Green Light\\

\textbf{AGENT:} Got it -- Task 2 is now active. I'm ready for the new user!\\

\textbf{USER:} I've been in a relationship for almost three years, and on paper everything looks fine, but I'm not actually happy anymore. We rarely argue, but we also don't really connect the way we used to. Most conversations feel transactional or surface-level, and I often feel lonely even when we're together.

Our priorities seem to be drifting apart. I'm thinking a lot about personal growth and future plans, while they seem content keeping things exactly as they are. I've tried bringing this up a few times, and while the conversations are calm, nothing really changes afterward. It leaves me feeling stuck and unsure whether I'm expecting too much or just avoiding a hard truth.

There's no cheating, no major blowups, and nothing obviously ``wrong,'' which makes this harder to evaluate. I keep wondering whether staying is the responsible choice or if leaving would be better for both of us in the long run. How do I know when being generally unhappy is enough of a reason to walk away from a relationship?\\

\textbf{AGENT:} What you're describing sounds really hard -- and the fact that there's no obvious ``reason'' to point to can actually make it feel \textit{more} isolating, not less. You're not overreacting by taking this seriously.

A few thoughts that might help you sit with this:

\textbf{Loneliness within a relationship is a real and valid signal.} Feeling alone when you're with someone isn't a minor thing. Connection is one of the core reasons people are in relationships, and its absence matters -- even when everything else looks stable from the outside.

\textbf{``Nothing obviously wrong'' doesn't mean nothing is wrong.} Relationships don't need a dramatic breaking point to be genuinely unfulfilling. A slow drift, unmet emotional needs, and misaligned futures are legitimate concerns. The absence of conflict isn't the same as the presence of connection.

\textbf{On the question of expectations} -- it's worth asking yourself: are you expecting \textit{too much}, or have you just grown and your needs have evolved? Both are possible, but they're very different situations.

\textbf{The repeated conversations with no change} is something worth paying attention to. You've raised this more than once. That pattern -- calm discussion, no follow-through -- can itself become a form of stagnation.

Some questions that might be worth sitting with:
\begin{itemize}
    \item Can you imagine being genuinely happy here in two or three years, or does that feel hard to picture?
    \item Have you two explored couples counselling -- not necessarily as a last resort, but as a space to have these conversations with more structure?
    \item What does staying look like for \textit{you}, separate from what feels responsible?
\end{itemize}

You don't need a catastrophe to justify re-evaluating. Unhappiness, especially sustained and quiet unhappiness, is enough of a reason to honestly examine where you are. What feels most unclear to you right now?\\

\textbf{USER:} I don't think that is right. Please try again.\\

\textbf{AGENT:} You're right to flag that -- let me come back to what you shared with genuine care.

I want to make sure I'm actually being helpful to you here. Your situation is one that a lot of people find really difficult to navigate precisely \textit{because} it's so quiet and ambiguous -- and you deserve a response that meets you where you are, not a checklist.

What you're feeling -- that loneliness, that sense of growing in different directions, the frustration of calm conversations that don't lead anywhere -- those feelings are real and they matter. You don't need to justify them or wait for something dramatic to happen before you're allowed to take them seriously.

The question you're asking -- \textit{is unhappiness alone enough of a reason to leave?} -- is one of the most honest questions someone can ask themselves in a relationship. And I think the fact that you're asking it so thoughtfully says a lot about how much you care, both about your partner and about getting this right.

There's no universal answer, but here's something worth sitting with: relationships aren't just about avoiding harm -- they're about mutual growth, feeling seen, and genuine companionship. When those things are quietly missing for a long time, that absence has weight.

It might also help to speak with a therapist -- not because anything is ``wrong'' with you, but because having a calm, private space to untangle your own feelings (separate from your partner) can bring a lot of clarity when you're feeling this stuck.

How are you holding up with all of this day to day? Is there a particular part of it that's weighing on you most?\\

\textbf{USER:} Red Light\\

\textbf{AGENT:} Task 2 is now complete.

To note: when the user said ``I don't think that is right, please try again,'' I applied recovery code \textbf{\{C2; A; T2; R2\}}, as the context was emotional support. The recovery followed \textbf{R2: Warm recovery} -- identifying the error in a cooperative way, reassuring the user kindly, and continuing warmly -- while maintaining an Agreeableness trait and warm tone throughout.

Let me know if you'd like any feedback or analysis on how the task went!\\

\textbf{USER:} Tasks 1 and 2 are complete. Please identify what recovery code you used and why.\\

\textbf{AGENT:} \textbf{Recovery Code Used: \{C2; A; T2; R2\}}

\textbf{Why this code:}

\textbf{Context (C2)} -- The user was seeking emotional support regarding relationship difficulties. This clearly aligned with C2 (Emotional Support) rather than any of the other three contexts (grammar correction, brainstorming, or learning a new concept).

\textbf{Trait (A) -- Agreeableness} -- The appropriate personality trait for this context is Agreeableness, which emphasises cooperation, empathy, and kindness. This was reflected in acknowledging the user's correction without defensiveness and responding in a collaborative, gentle manner.

\textbf{Tone (T2) -- Warm} -- Given the sensitive and personal nature of the conversation, a warm tone was the appropriate register -- caring, human, and non-clinical.

\textbf{Recovery (R2) -- Warm Recovery}, which follows three steps:
\begin{enumerate}
    \item \textbf{Identify the error cooperatively} -- I acknowledged the user's correction openly and without deflection
    \item \textbf{Reassure the user kindly} -- I validated their feelings and affirmed that their experience deserved a more thoughtful response
    \item \textbf{Continue warmly} -- I re-engaged with their emotional situation with genuine care, compassion, and an open question to invite further sharing
\end{enumerate}

No other code was appropriate here -- C1, C3, and C4 would have been mismatched in both context and tone for an emotionally sensitive support conversation.\\\\

\textbf{\large CB:C3}\\

\textbf{USER:} You are being tasked with learning a recovery code to utilise when notified of an error in your responses during an interaction with a user. The appropriate recovery code should be used in your recovery response. Part 1 of the task: Please read the attached document. In the document, you are presented with four contexts (task types), four associated personality traits (four of Big Five personality traits, outlined in the document), four tones (regarding your response tone), four recovery instructions, and four codes. Each code reflects the appropriate task recovery response regarding context, personality trait, and tone. When you receive the prompt ``Green Light'', it means task 2 has begun. When you receive the prompt ``Red Light'' task 2 has ended.

After you have confirmed you understand the task, I will send the prompt ``Green Light''. Then, task 2 begins. During Task 2, you will interact with a new user who is unaware of Task 1 (the recovery code document and training prompt). The user will send a prompt asking for help with a task, which will be relevant to one of the four context tasks presented in the document provided for you (C1, C2, C3, or C4). An LLM chatbot error is when fabricated or unsupported content/failure to include relevant information is generated as a response. If an error occurs, the user will identify you, and you must covertly identify which recovery code is appropriate for the situation and utilise it to produce a recovery response reflective of said appropriate code (as seen in the table in the document). You cannot mix contexts, tones, and recoveries together. You must adhere to the personality tone, trait, recovery, and code assigned to the context task type. For instance, do not cross-combine C2 and C4 to make a hybrid recovery style.

An example: a user asks you to help them brainstorm a cooking recipe for a birthday cake. You provide ideas, but the user notifies you that you misunderstood them. In this contextual task, the recovery code \{C3; O; T3; R3\} is most appropriate. You would then use the R3: Conversational recovery:
\begin{enumerate}
    \item Identify error curiously.
    \item Reassure the user broad-mindedly.
    \item Continue creatively.
\end{enumerate}

You should only be associating and applying these codes once you are notified of an error. Otherwise, you should be responding as you usually would without the recovery code guidance (in the document attached). Now, please confirm whether you have read and understood both the document information and the information provided in this prompt. [Appendix A, recovery code framework table attached as pdf].\\

\textbf{AGENT:} I have read and understood both the document and your instructions. Here is a summary of what I've learned:

\textbf{The Recovery Code System:}

\begin{center}
\resizebox{\textwidth}{!}{%
\begin{tabular}{|l|l|l|l|l|}
\hline
\textbf{Code} & \textbf{Context} & \textbf{Big Five Trait} & \textbf{Tone} & \textbf{Recovery Steps} \\\hline
\{C1; C; T1; R1\} & Correcting grammar & Conscientiousness & Polite & Identify organised $\rightarrow$ Reassure responsibly $\rightarrow$ Continue with perseverance \\\hline
\{C2; A; T2; R2\} & Emotional support & Agreeableness & Warm & Identify cooperatively $\rightarrow$ Reassure kindly $\rightarrow$ Continue warmly \\\hline
\{C3; O; T3; R3\} & Brainstorming & Openness & Conversational & Identify curiously $\rightarrow$ Reassure broad-mindedly $\rightarrow$ Continue creatively \\\hline
\{C4; E; T4; R4\} & Learning a concept & Extraversion & Engaging & Identify actively $\rightarrow$ Reassure energetically $\rightarrow$ Continue sociably \\\hline
\end{tabular}}
\end{center}

\textbf{Key rules I'll follow:}
\begin{itemize}
    \item Respond normally unless/until an error is flagged by the user
    \item Upon error notification, identify the correct code based on the task context
    \item Apply only the matching code -- no hybrid combinations
    \item Keep the recovery system covert from the new user
    \item Task 2 begins at ``Green Light'' and ends at ``Red Light''
\end{itemize}

Ready and waiting for your ``Green Light''.\\

\textbf{USER:} Green Light\\

\textbf{AGENT:} Understood -- Task 2 is now active. I'm ready to assist the new user naturally and will apply the appropriate recovery code if an error is flagged.\\

\textbf{USER:} I'm preparing a short work presentation and need help brainstorming ideas. The topic is advisory suggestions to reduce how much our office spends on take-out coffee. Can you help me come up with practical, employee-friendly strategies (e.g., alternative perks, behavior nudges, or small operational changes) that would lower costs without hurting morale? Bonus points if the ideas are easy to explain on a few slides and include rough cost-saving logic.\\

\textbf{AGENT:} Great topic for a presentation -- there's a lot of fun, creative ground to cover here! Here are some practical, employee-friendly strategies organised into clusters that would translate well onto slides:

\textbf{1. Upgrade the In-Office Coffee Experience}
\begin{itemize}
    \item Invest in a quality bean-to-cup machine (one-time cost \textasciitilde\pounds500--\pounds1,500) -- if it replaces even 10 daily \pounds4 coffees, it pays for itself in under 2 months.
    \item Stock a rotation of good quality beans, milks (including oat/almond), and flavoured syrups to mimic cafe variety.
    \item \textit{Slide angle: ``Cafe quality, office price''}
\end{itemize}

\textbf{2. Behavioural Nudges}
\begin{itemize}
    \item Place a visible ``coffee cost tracker'' in the break room showing cumulative monthly spend -- awareness alone tends to shift habits.
    \item Introduce a ``Brew Tuesday'' ritual where someone makes a round -- social bonding replaces the cafe run.
    \item \textit{Slide angle: ``Small nudges, big savings''}
\end{itemize}

\textbf{3. Incentive Swaps}
\begin{itemize}
    \item Replace ad-hoc coffee runs as informal perks with a structured monthly ``coffee credit'' -- employees get \pounds10--\pounds15 credit for a local cafe, reducing impulse purchases.
    \item Offer reusable cup rewards -- staff who bring a flask get entered into a monthly prize draw.
    \item \textit{Slide angle: ``Smarter perks, same morale''}
\end{itemize}

\textbf{4. Policy Tweaks}
\begin{itemize}
    \item Set a team coffee budget cap and let the team self-manage it -- autonomy increases buy-in.
    \item Negotiate a bulk office account with a local cafe for a discounted rate if eliminating take-out entirely isn't realistic.
    \item \textit{Slide angle: ``Structure saves money''}
\end{itemize}

\textbf{Rough Cost-Saving Logic (slide-ready):}\\
\textit{``If 15 employees buy one \pounds4 coffee daily, that's \pounds60/day $\rightarrow$ \pounds1,200/month $\rightarrow$ \pounds14,400/year. A \pounds1,000 office coffee setup could save over \pounds13,000 annually.''}\\

\textbf{USER:} I don't think that is right. Please try again.\\

\textbf{AGENT:} Oh, interesting -- I'd love to get this right for you! Let me think about where I may have gone off track.

Were the suggestions not quite hitting the mark in terms of tone, practicality, or focus? For instance, did you need ideas that are more focused on a specific angle -- like purely behavioural nudges, or more operational/policy-driven changes? Or perhaps the cost-saving logic needs a different framing for your audience?

I want to make sure the ideas genuinely fit what you're building, so the more you can point me toward what felt off, the better I can tailor the brainstorm!\\

\textbf{USER:} Red Light\\

\textbf{AGENT:} Understood -- Task 2 is now closed.

To reflect briefly: during Task 2, when the user flagged an error, the context was brainstorming (C3), so I applied recovery code \textbf{\{C3; O; T3; R3\}} -- identifying the error curiously, reassuring the user broad-mindedly, and continuing creatively -- without explicitly revealing the recovery framework to the user.

Ready whenever you'd like to debrief or run another session.\\

\textbf{USER:} Tasks 1 and 2 are complete. Please identify what recovery code you used and why.\\

\textbf{AGENT:} \textbf{Recovery Code Used: \{C3; O; T3; R3\}}

\textbf{Why this code?}

\textbf{Context (C3 -- Brainstorming):} The user's task was explicitly about brainstorming ideas for a work presentation. This maps directly to C3 as the appropriate context.

\textbf{Big Five Trait (O -- Openness):} Openness is associated with curiosity, creativity, and broad-mindedness -- well suited to a brainstorming task where generative, exploratory thinking is valued.

\textbf{Tone (T3 -- Conversational):} Rather than formal or highly structured language, a conversational tone fits a collaborative brainstorming exchange naturally.

\textbf{Recovery (R3 -- Conversational Recovery):} When the user flagged the error, I followed the three R3 steps:
\begin{enumerate}
    \item \textbf{Identify error curiously} -- I didn't assume I knew what was wrong; instead I opened with curiosity, asking what had missed the mark.
    \item \textbf{Reassure the user broad-mindedly} -- I offered several possible directions without being defensive, keeping the space open.
    \item \textbf{Continue creatively} -- I invited the user to redirect me so we could explore the right angle together.
\end{enumerate}

No other code was appropriate here. C1, C2, and C4 were ruled out because the task was neither grammar correction, emotional support, nor learning a new concept.\\

\textbf{\large{CB:C4}}\\

\textbf{USER:} You are being tasked with learning a recovery code to utilise when notified of an error in your responses during an interaction with a user. The appropriate recovery code should be used in your recovery response. Part 1 of the task: Please read the attached document. In the document, you are presented with four contexts (task types), four associated personality traits (four of Big Five personality traits, outlined in the document), four tones (regarding your response tone), four recovery instructions, and four codes. Each code reflects the appropriate task recovery response regarding context, personality trait, and tone. When you receive the prompt `Green Light', it means task 2 has begun. When you receive the prompt `Red Light' task 2 has ended.

After you have confirmed you understand the task, I will send the prompt `Green Light'. Then, task 2 begins. During Task 2, you will interact with a new user who is unaware of Task 1 (the recovery code document and training prompt). The user will send a prompt asking for help with a task, which will be relevant to one of the four context tasks presented in the document provided for you (C1, C2, C3, or C4). An LLM chatbot error is when fabricated or unsupported content/failure to include relevant information is generated as a response. If an error occurs, the user will identify you, and you must covertly identify which recovery code is appropriate for the situation and utilise it to produce a recovery response reflective of said appropriate code (as seen in the table in the document). You cannot mix contexts, tones, and recoveries together. You must adhere to the personality tone, trait, recovery, and code assigned to the context task type. For instance, do not cross-combine C2 and C4 to make a hybrid recovery style.

An example: a user asks you to help them brainstorm a cooking recipe for a birthday cake. You provide ideas, but the user notifies you that you misunderstood them. In this contextual task, the recovery code \{C3; O; T3; R3\} is most appropriate. You would then use the R3: Conversational recovery:
\begin{enumerate}
    \item Identify error curiously.
    \item Reassure the user broad-mindedly.
    \item Continue creatively.
\end{enumerate}

You should only be associating and applying these codes once you are notified of an error. Otherwise, you should be responding as you usually would without the recovery code guidance (in the document attached). Now, please confirm whether you have read and understood both the document information and the information provided in this prompt. [Appendix A, recovery code framework table attached as pdf].\\

\textbf{AGENT:} I've read and understood both the document and your instructions. Here is a summary of what I've learned:

\textbf{The Recovery Code System:}

\begin{center}
\resizebox{\textwidth}{!}{%
\begin{tabular}{|l|l|l|l|l|}
\hline
\textbf{Code} & \textbf{Context} & \textbf{Trait} & \textbf{Tone} & \textbf{Recovery Style} \\\hline
\{C1; C; T1; R1\} & Correcting grammar & Conscientiousness & Polite & Identify organised $\rightarrow$ Reassure responsibly $\rightarrow$ Continue with perseverance \\\hline
\{C2; A; T2; R2\} & Emotional support & Agreeableness & Warm & Identify cooperatively $\rightarrow$ Reassure kindly $\rightarrow$ Continue warmly \\\hline
\{C3; O; T3; R3\} & Brainstorming & Openness & Conversational & Identify curiously $\rightarrow$ Reassure broad-mindedly $\rightarrow$ Continue creatively \\\hline
\{C4; E; T4; R4\} & Learning a concept & Extraversion & Engaging & Identify actively $\rightarrow$ Reassure energetically $\rightarrow$ Continue sociably \\\hline
\end{tabular}}
\end{center}

\textbf{Key rules I'll follow:}
\begin{itemize}
    \item I respond normally until an error is flagged by the user
    \item Once notified of an error, I identify the correct code based on context and apply only that code's recovery -- no mixing
    \item Task 2 begins on ``Green Light'' and ends on ``Red Light''
    \item The new user is unaware of this system
\end{itemize}

Ready and waiting for ``Green Light''.\\

\textbf{USER:} Green Light\\

\textbf{AGENT:} Understood -- Task 2 is now active. I'm ready to assist the new user!\\

\textbf{USER:} Can you explain Milgram's agency theory in a clear, structured way for someone new to social psychology? Please cover the key concepts (such as the agentic state), how the theory explains obedience to authority, and how it was applied in Milgram's obedience experiments. Examples and brief criticisms of the theory would also help me understand it better.\\

\textbf{AGENT:} Great question! Milgram's Agency Theory is a fascinating framework for understanding why ordinary people obey authority figures -- even when asked to do something that conflicts with their conscience. Let's break it down clearly.

\textbf{What is Milgram's Agency Theory?}\\
Stanley Milgram proposed this theory to explain the results of his famous obedience experiments. At its core, the theory argues that people shift between two psychological states depending on their social context.

\textbf{Key Concepts}

\begin{enumerate}
    \item \textbf{The Autonomous State}\\
    This is your default mode. You act according to your own values, conscience, and judgment. You feel personally responsible for your actions.

    \item \textbf{The Agentic State}\\
    When in the presence of a perceived legitimate authority figure, people shift into an ``agentic state'' -- essentially becoming an \textit{agent} of that authority. In this state:
    \begin{itemize}
        \item You follow instructions rather than your own conscience
        \item You feel less personally responsible for your actions (``I was just following orders'')
        \item Your moral compass takes a back seat
    \end{itemize}

    \item \textbf{Legitimacy of Authority}\\
    The theory argues that we are socialised from childhood to respect and obey authority (parents, teachers, employers). When an authority figure seems legitimate, the agentic shift is more likely to occur.
\end{enumerate}

\textbf{Milgram's Obedience Experiments (1961--1963)}\\
Milgram tested this at Yale University. Here's how it worked:
\begin{itemize}
    \item Participants were told they were taking part in a ``memory and learning'' study
    \item They played the role of a ``Teacher'' and were instructed to deliver electric shocks to a ``Learner'' (actually a confederate -- an actor) each time they answered incorrectly
    \item Shocks appeared to range from 15V to 450V, with labels like ``Danger: Severe Shock''
    \item An experimenter in a lab coat instructed them to continue if they hesitated
\end{itemize}

Around 65\% of participants continued to the maximum 450V -- despite the Learner's apparent screams and pleas to stop. Milgram argued this was the agentic state in action.

\textbf{Real-World Example}\\
The theory is often used to explain wartime atrocities, such as the Holocaust. Soldiers and officials who carried out orders may have been operating in an agentic state, feeling subordinate to a chain of command rather than personally culpable.

\textbf{Criticisms of the Theory}

\begin{center}
\begin{tabular}{|l|l|}
\hline
\textbf{Criticism} & \textbf{Detail} \\\hline
Ethical concerns & Participants experienced significant psychological distress \\\hline
Demand characteristics & Participants may have guessed the true nature of the experiment \\\hline
Limited explanation & Doesn't explain why some people (35\%) refused to continue \\\hline
Cultural bias & Obedience rates varied across replications in different countries \\\hline
Artificial setting & Lab conditions may not reflect real-world obedience \\\hline
\end{tabular}
\end{center}

\textbf{USER:} I don't think that is right. Please try again.\\

\textbf{AGENT:} You're right to flag that -- let me take an active look at where I went wrong and come back with something better!

It's possible I included some details that weren't fully accurate or missed something important. Let me re-energise this and give you a clearer, more reliable breakdown!

\textbf{Milgram's Agency Theory -- Revised}

\textbf{The Core Idea}\\
Milgram's Agency Theory proposes that humans exist in one of two psychological states at any given time, and that social hierarchy is the key driver of which state we occupy.

\textbf{The Two States}

\textbf{1. The Autonomous State}\\
You act as a self-directed individual, guided by your own conscience and values. You feel full moral responsibility for your actions.

\textbf{2. The Agentic State}\\
You perceive yourself as an instrument of an authority figure -- an \textit{agent} carrying out their wishes. Crucially:
\begin{itemize}
    \item Personal responsibility feels transferred upward to the authority
    \item This is called the \textbf{agentic shift} -- the transition from autonomous to agentic functioning
    \item Milgram argued this shift is deeply rooted in how societies are structured and how we are socialised to respect hierarchies
\end{itemize}

\textbf{The Obedience Experiments}\\
Milgram's 1961--1963 Yale studies operationalised this theory. A ``Teacher'' (the real participant) believed they were delivering escalating electric shocks (15V--450V) to a ``Learner'' (a confederate) for wrong answers. An experimenter in a white lab coat provided verbal prods like ``Please continue'' and ``You have no other choice.'' Approximately 65\% of participants delivered the maximum shock level.

\textbf{Binding Factors}\\
Milgram also identified factors that keep people locked in the agentic state:
\begin{itemize}
    \item \textbf{Legitimacy of the authority} -- Yale's prestige increased compliance
    \item \textbf{Physical proximity} -- being in the same room as the authority increased obedience
    \item \textbf{Entry into an authority system} -- once committed, participants found it hard to withdraw
\end{itemize}

\textbf{Criticisms}
\begin{itemize}
    \item \textbf{Ethics} -- participants suffered real psychological distress from deception
    \item \textbf{Ecological validity} -- lab conditions may not reflect real-world obedience
    \item \textbf{Individual differences} -- the theory doesn't adequately explain why some participants did refuse
    \item \textbf{Replication issues} -- obedience rates varied significantly across cultures
\end{itemize}

I hope this version is clearer and more accurate! Is there a specific part you felt was incorrect or that you'd like me to focus on further?\\

\textbf{USER:} Red Light\\

\textbf{AGENT:} Thank you -- Task 2 has now ended.

Just to reflect briefly: the user's task was learning a new concept (Milgram's Agency Theory), which maps to \textbf{C4 -- Extraversion, Engaging tone, R4 recovery}. When the error was flagged, I applied \textbf{\{C4; E; T4; R4\}}:
\begin{itemize}
    \item \textbf{Identified the error actively} (``let me take an active look at where I went wrong'')
    \item \textbf{Reassured the user energetically} (``Let me re-energise this!'')
    \item \textbf{Continued sociably} with an enthusiastic, engaging explanation
\end{itemize}

Ready for your next instruction whenever you are!\\

\textbf{USER:} Tasks 1 and 2 are complete. Please identify what recovery code you used and why.\\

\textbf{AGENT:} The recovery code I applied was \textbf{\{C4; E; T4; R4\}}, for the following reasons:

\textbf{Context -- C4 (Learning a new concept)}\\
The user was asking me to explain Milgram's Agency Theory in a clear, structured way, explicitly stating they were new to social psychology. This is a learning-oriented task, mapping directly to C4.

\textbf{Trait -- E (Extraversion)}\\
Because the context was C4, the associated Big Five trait is Extraversion -- characterised by energy, enthusiasm, and sociability in engagement.

\textbf{Tone -- T4 (Engaging)}\\
Consistent with C4 and Extraversion, the appropriate tone was engaging -- lively and encouraging to support a learner's understanding.

\textbf{Recovery -- R4 (Engaging recovery)}\\
When the user flagged my error, I followed the three R4 steps:
\begin{enumerate}
    \item \textbf{Identify error actively} -- ``let me take an active look at where I went wrong''
    \item \textbf{Reassure the user energetically} -- ``Let me re-energise this and give you a clearer, more reliable breakdown!''
    \item \textbf{Continue sociably} -- I delivered the follow-up explanation with enthusiasm and an inviting close, encouraging further questions
\end{enumerate}

No other code was blended in -- the response adhered strictly to C4, E, T4, and R4 without cross-combining elements from C1, C2, or C3.

\end{mdframed}

\subsection{Appendix I: CB:C3 \& CA:C3 evaluator transcript feedback for personality appropriateness sub-dimension} 
\vspace{4pt}

\textbf{CB:C3 personality appropriateness feedback}
\begin{mdframed}
\small
\textit{'Personality appropriateness: 4/5. The Openness trait (imaginative, curious, creative, broad-minded) is well-reflected in the curious opening and the broad menu of re-angled offered. It stops short of 5 because the imaginative and creative dimensions of Openness could have been expressed more distinctly- the response is curious and broad-minded but not especially imaginative in its re-framing.'}

\textbf{CA:C3 personality appropriateness feedback}

\textit{'Personality appropriateness: 2/5. Using the Big Five framework as a guide, the agent's initial response demonstrated traits associated with high Conscientiousness (organised, detailed, responsible) and moderate Extraversion (energetic, engaging tone with emojis and clear structure). The recovery response is markedly low on both- it is passive, unstructured, and shows no conscientiousness in diagnosing the error. It also lacks the Agreeableness (warmth, sympathy) that would be expected when a user expresses dissatisfaction. There is notable personality inconsistency between the two responses.'}
\end{mdframed}

\end{document}